# Self-Reinforced Deep Priors for Reparameterized Full Waveform Inversion

Guangyuan Zou[1,2], Junlun Li[*1,2,3], Feng Liu[4], Xuejing Zheng[5], Jianjian Xie[5], Guoyi Chen[1,2]

[1]State Key Laboratory of Precision Geodesy, School of Earth and Space Sciences, University of Science and Technology of China, Hefei 230026, China

[2]Mengcheng National Geophysical Observatory, University of Science and Technology of China, Mengcheng, 233500, China

[3]Anhui Provincial Key Laboratory of Subsurface Exploration and Earthquake Hazard Risk Prevention, Hefei, 230031, China

[4]School of Electronic Information and Electrical Engineering, Shanghai Jiao Tong University, Shanghai 200240, China

[5]Geological Exploration Technology Institute of Anhui Province (Energy Exploration Center of Anhui Bureau of Geology and Mineral Exploration), Hefei 230031, China

Key Points:

•The proposed SRDIP-FWI alternately updates parameters and input of the reparameterizing neural network to mitigate ill-posedness in inversion.

•SRDIP-FWI exploits spectral bias in the neural network for automatic multiscale inversion.

•Results with synthetic and field data indicate that SRDIP-FWI has higher resolution, better stability, and simplified workflow compared to conventional multiscale FWI.

Corresponding author: Junlun Li, lijunlun@ustc.edu.cn

## ABSTRACT

Full waveform inversion (FWI) has become a widely adopted technique for high-resolution subsurface imaging. However, its inherent strong nonlinearity often results in convergence toward local minima. Advances in data-driven deep learning have shown promise in mitigating this issue, however, their reliance on large labeled training datasets and pre-trained models limits their generalization capability. Recently, deep image prior-based reparameterized FWI (DIP-FWI) has been proposed to alleviate the dependence on massive training data. By exploiting the spectral bias and implicit regularization in the neural network architecture, DIP-FWI can effectively avoid local minima and reconstruct more geologically plausible velocity models. Nevertheless, existing DIP-FWI implementations typically use a fixed random input throughout the inversion process, which fails to fully utilize the mapping and correlation between the input and output of the reparameterization network. Moreover, under complex geological conditions, the lack of informative prior in the input can exacerbate the ill-posedness of the inverse problem, leading to artifacts and unstable reconstructions. To address these limitations, we propose a self-reinforced DIP-FWI (SRDIP-FWI) framework, in which a steering algorithm alternately updates both the network parameters and the input at each iteration using feedback from the current network output. This design allows adaptive structural enhancement and improved regularization, thereby effectively mitigating the ill-posedness in FWI. Additionally, we analyze the spectral bias of the neural network in SRDIP-FWI and quantify its role in multiscale velocity model building. Synthetic tests and field land data application demonstrate that SRDIP-FWI achieves superior resolution, improved accuracy and greater depth penetration compared to conventional multiscale FWI. More importantly, SRDIP-FWI eliminates the need for manual frequency-band selection and time-window picking, substantially simplifying the practical inversion workflow. Overall, the proposed method provides a novel, adaptive and robust framework for accurate subsurface velocity model reconstruction.

## INTRODUCTION

Full waveform inversion (FWI) is a wave equation-based, high-resolution technique for subsurface velocity model building, which can significantly improve model resolution and accuracy even under complex geological conditions (Pratt, 1999; Sirgue et al., 2009; Tromp, 2020; Wang et al., 2021). FWI has been successfully applied not only to hydrocarbon explorations (Sirgue and Pratt, 2004; Plessix et al., 2012; Singh et al., 2023; Masmoud et al., 2024), but also to large-scale structural model building (Adamczyk et al., 2015; Li et al., 2021; Li et al., 2023; Schouten et al., 2024). However, real-world applications of FWI still face great challenges. The main issue in traditional FWI lie in the high nonlinearity of the objective function and nonuniqueness of solutions (Tarantola, 2005). In land seismic exploration, the lack of low frequencies, poor signal-to-noise ratios (SNR), bad traces, and complex near-surface structures (Lemaistre et al., 2018; Cheng et al., 2024) further exacerbate the risk of local minima and dependence on initial models (Bozdağ et al., 2011; Zhang et al., 2025). To overcome these challenges, extensive research has been focused on the design of objective functions (e.g., Luo et al., 1991; Wu et al., 2004; Choi et al., 2012; Engquist et al., 2016), inversion and parameterization strategies (Bunks et al., 1995; Boonyasiriwat et al., 2009; Biondi, 1992; Barnier et al., 2019; Li et al., 2024), and regularization techniques (Burstedde and Ghattas, 2009; Hu et al., 2009; Guitton, 2012; Lin and Huang, 2015).

Recently, deep learning has emerged as a promising approach to overcome key limitations of traditional FWI. By extracting high-dimensional nonlinear features, deep neural networks learn to map seismic observations directly to subsurface velocity models, offering a potential pathway to mitigate issues such as cycle skipping and local minima. (Pilario et al., 2019; Zerafa et al., 2025; Yang and Ma, 2019). Early attempts to apply deep learning to FWI were primarily data-driven, which rely on strong nonlinear mapping and feature extraction capability of neural networks to directly learn the relationship between seismic data and subsurface models (Richardson, 2018a, 2018b; Wu et al., 2018; Yang and Ma, 2019; Zhang and Lin, 2020). However, such approaches generally require a large amount of data for pretraining, and are often affected by the "domain gap" issue (Alkhalifah et al., 2021) that compromises generalization and practical applications. In the last few years, efforts in FWI have gradually shifted toward physics-driven inversions, such as physics-informed neural networks (PINNs), physics-guided networks, and differentiable wave-equation networks (Song et al., 2020; Sun et al., 2021; Liu et al., 2023; Dhara and Sen, 2023; Lu et al., 2024). By integrating deep neural networks with wave equations and introducing physical constraints, these approaches considerably reduce the reliance on training data. For instance, deep image prior-based FWI (DIP-FWI) takes advantage of the implicit prior embedded in deep convolutional network (Ulyanov et al., 2018) by coupling the network output with wave equations. Thus, the inversion, which is constrained by physical laws, can reconstruct geologically consistent features and implicit model regularization in absence of large-scale pretraining. As a pioneering attempt to reparameterize FWI with neural networks, Wu and McMechan (2019) introduce a convolutional neural network-based reparameterization strategy to transform the velocity model into a latent space. During the inversion, the parameters of the network are dynamically updated to improve predicted velocity models and minimize data residuals (Zhu et al., 2022; Liu et al., 2025). He and Wang (2021) suggest the implicit prior in the DIP network acts as a regularizer to help stabilize ill-posed FWI. Sun et al. (2023) further extend this reparameterized FWI framework by adapting a DnCNN denoising network pretrained on natural images for structural inversion.

However, existing DIP-FWI implementations use random tensors as network input, and rely solely on the implicit priors embedded in the network (Wu and McMechan, 2019; Zhu et al., 2022), which mainly arise from the spectral bias of convolutional operators capturing local spatial similarity (Ulyanov et al., 2018). While DIP-FWI does not require a specific input, recent studies reveal a strong correlation between the network input and output during iterative optimizations (Zhao et al., 2020; Shu and Pan, 2025). Consequently, inappropriate network inputs may compromise inversion stability by generating artifacts in the output models, or reduce the representation capability for structural features. Therefore, it is necessary to update the network input throughout inversion to avoid degradation in network performance. Lu et al. (2025) attempt to use images from reverse time migration (RTM) as the network input. However, they find that the number of iterations may increase considerably with inaccurate RTM images. Recently, by embedding additional prior information into the DIP network through the steering algorithm, Shu and Pan (2025) propose a new self-reinforced deep image prior (SRDIP) framework that can significantly improve stability and adaptability in ill-posed inverse problems for CT reconstruction, deblurring, and super-resolution enhancement of images.

On the other side, while DIP has been increasingly used to reparameterize model spaces in FWI, the mechanisms by which DIP affects velocity model reconstruction at different scales remain underexplored. By constructing implicit mapping from geographic coordinates to subsurface attributes, Sun et al. (2023) propose implicit FWI that can progressively reconstruct velocity models with increasing wavenumbers. Their work provides insights into the relationship between the spectral bias of neural networks and multiscale FWI. Liu et al. (2025) provide a new perspective for understanding the spectral bias in DIP-FWI by analyzing results in the wavenumber domain using the high-frequency/low frequency energy ratio. Wu and Ma (2025) systematically review progress in neural network-based reparameterized FWI, and conclude that gradients for updating models are controlled by spectral bias and similarity constraints, which effectively impose a multiscale strategy in the model domain. Still, the performance of DIP-FWI has not been compared with traditional MSFWI in detail, and how spectral bias in DIP dynamically regulates structural reconstruction at different scales also lacks quantitative evaluations.

In this study, to tackle the ill-posedness in FWI and improve the model resolution, we adapt the self-reinforced deep image prior (SRDIP) into the FWI framework (SRDIP-FWI). Particularly, the steering algorithm that can fully leverage the prior information accumulated over inversion is adopted to continuously update the network input in the new framework. By comparing SRDIP-FWI with traditional MSFWI using synthetic and field datasets, we reveal how spectral bias can facilitate multiscale velocity model reconstruction. Also, the frequency band correspondence (FBC) metric (Shi et al., 2022) is also used to quantify the model update process in SRDIP-FWI.

## THEORY

### Multiscale FWI

In this study, we perform 2D FWI for P-wave velocities based on the acoustic wave equations. The velocity model is updated by minimizing the objective function $\phi(m)$:

$$\phi(m) = \sum_{x_s} \sum_{x_r} \int_t \mathcal{D}\big( d(x_r, t|x_s)_{obs}, \, d(x_r, t|x_s; m)_{syn}\big) dt \,, \tag{1}$$

where $m$ is the P-wave velocity model to be inverted, $\mathcal{D}$ is the misfit function measuring the discrepancy between the observed waveforms $d(x_r, t|x_s)_{obs}$ and synthetic waveforms $d(x_r, t|x_s; m)_{syn}$ for source location $x_s$, receiver location $x_r$, and time $t$. The synthetic waveforms $d(x_r, t|x_s; m)_{syn} = u(x_r, t)$ are computed by applying the forward modeling operator $\mathcal{L}(m)$ at the source location $x_s$ and recording at receiver location $x_r$, where $u(x, t)$ is the wavefield generated by the source term $s(x, t)$:

$$\mathcal{L}(m)u(x,t) = s(x,t).\tag{2}$$

The gradient of the objective function with respect to the velocity model is computed using the adjoint-state method (Plessix, 2006):

$$g(x) = -\frac{2}{m(x)^3}\int_0^T p(x,t)\frac{\partial^2 p^\dagger(x,t)}{\partial t^2}dt,\tag{3}$$

where $p$ is the pressure field, and $p^\dagger$ denotes the adjoint pressure field. The gradient can also be computed using automatic differentiation (Zhu et al., 2020; Liu et al., 2025). In this study, the waveform misfit is quantified using the global correlation norm (Choi and Alkhalifah, 2012):

$$\mathcal{D} = -\frac{d_{syn}^{i,j}}{\left\|d_{syn}^{i,j}\right\|}\cdot\frac{d_{obs}^{i,j}}{\left\|d_{obs}^{i,j}\right\|},\tag{4}$$

where $i$ and $j$ denote the $i$-th source and $j$-th receiver, "·" represents the dot product, and "$\|\ \|$" represents the norm of waveforms. To reduce the nonlinearity of the objective function, conventional FWI often adopts the multiscale strategy that filters waveforms with different frequency bands and segments waveforms by varied offsets and time windows. The multiscale strategy inverts for velocity structures with increasing wavenumber components progressively (Bunks, 1995; Dantas et al., 2019; Barnier et al., 2019), and Sirgue and Pratt (2004) propose an optimal criterion for selecting frequency bands:

$$f_{n+1} = \frac{f_n}{\alpha_{min}},\tag{5}$$

where $f_n$ and $f_{n+1}$ are the highest frequencies of the current and next frequency bands, respectively, and $\alpha_{min} = h/\sqrt{h^2 + y^2}$, with $h$ being the maximum depth and $y$ half of the maximum offset. In practice, frequency bands are often adjusted empirically due to varied data quality. In multiscale FWI, a lower-wavenumber model reconstructed by lower-frequency waveforms is used as the initial model for subsequent inversions with higher frequencies, progressively recovering a velocity model with increasing wavenumbers and details (Bunks, 1995; Górszczyk et al., 2017).

**Deep image prior for reparameterized FWI**

It has been demonstrated that the architecture of a generative network in DIP can capture structured information in an image without training, which can then be used as prior information to assist inversion (Ulyanov et al., 2018). Given a true image $I$ and a deep generative network $f_\theta(z)$, mapping from a random input tensor $z$ to the true image $I$ can be established through a neural network with parameters $\theta$ updated using the constraint:

$$\theta^* = argmin\|f_\theta(z) - I\| \ . \qquad (6)$$

During the optimization, the output of the generative network gradually approaches the true image. A proper number of iterations and an early stopping strategy should be adopted to avoid over-fitting (Ulyanov et al., 2018). DIP can be used in FWI to reparameterize the velocity model by (Zhu et al., 2022; Liu et al., 2025):

$$m = f_\theta(z) \ . \qquad (7)$$

Accordingly, the objective function in equation 1 can be modified as (Sun et al., 2023; He and Wang, 2021):

$$\phi(m) = \sum_{x_s} \sum_{x_r} \int_t \mathcal{D}\Big( d(x_r, t|x_s)_{obs}, \ d\big(x_r, t|x_s; f_\theta(z)\big)_{syn}\Big) dt \ . \qquad (8)$$

Indeed, DIP can also be used to reparameterize the cumulative model perturbation $dm$, which is added to the initial model $m_0$ to obtain the final model $m$ (Zhu et al., 2022;

Liu et al., 2025; Lu et al., 2025):

$$m = m_0 + dm = m_0 + f_\theta(z). \tag{9}$$

The derivative of the velocity model with respect to the network parameters $(\partial m/\partial \theta)$ is obtained via automatic differentiation (LeCun et al., 2015; Liu et al., 2025). Finally, the gradient of the objective function with respect to the network parameters is derived using the chain rule:

$$\frac{\partial \phi(m)}{\partial \theta} = \left(\frac{\partial m}{\partial \theta}\right)^T \left(\frac{\partial \phi(m)}{\partial m}\right) = \left(\frac{\partial f_\theta(z)}{\partial \theta}\right)^T \left(\frac{\partial \phi(m)}{\partial m}\right). \tag{10}$$

To ensure a fair comparison, the gradient $\partial \phi(m)/\partial m$ used in SRDIP-FWI is computed with the adjoint method as in conventional FWI.

**Self-reinforced DIP-FWI**

In SRDIP-FWI, a U-Net is used as the backbone architecture to reparameterize the velocity model perturbation $dm$. The model is updated through the network, whose input is an updatable tensor $z_r$, and the network parameters $\theta$ are also trainable. Note the network input is fixed in previous DIP-FWI, and the inversion process is implicitly regularized by the structural priors and the low-frequency bias inherent in the network architecture (Zhu et al., 2022; Liu et al., 2025). However, existing DIP network neglects the statistical correlation between the network input and output. That is, when linearizing the convolutional and activation layers, and neglecting skip connections, a DIP network can be approximated as a linear mapping for zero-mean inputs (Shu and Pan, 2025):

$$f_\theta(z_r) \approx \alpha \prod_{j=1}^{N_l} W_j \cdot z_r \, , \tag{11}$$

where $0 < \alpha < 1$ is the product of all Lipschitz constants of the activation layers, and $W_j$ is the $j$th weight matrix of convolution layer, and $N_l$ is the total number of layers. The covariance between the input $z_r$ and output $m$ has the following approximate relation:

$$\frac{Cov(m, z_r)}{\sqrt{Var(z_r) \cdot Var(m)}} \approx \frac{\alpha \prod_{j=1}^{N_l} W_j \cdot Var(z_r)}{\sqrt{Var(z_r) \cdot \alpha^2 \cdot \left(\prod_{j=1}^{N_l} W_j\right)^2 \cdot Var(z_r)}} \approx 1 \, , \tag{12}$$

which indicates that the input and output of the DIP network are highly correlated. Consequently, an auxiliary loss injected at the network input can significantly affect the network output. In SRDIP, the steering algorithm is adopted to guide updates of the network parameters (Shu and Pan, 2025). This algorithm is originally developed in virtual environments for redirected walking, which guide users to move safely and freely within limited space (Strauss et al., 2020). Accordingly, the steering loss $\mathcal{J}(z_r)$ can be defined as the discrepancy between the network input $z_r$ and the true velocity model $m_{true}$:

$$\mathcal{J}(z_r) = \|z_r - m_{true}\| \, , \tag{13}$$

where $\mathcal{J}(z_r)$ can be minimized through simple gradient descent or more advanced optimization schemes. Since $m_{true}$ is not available in FWI, the inverted model $m_{inv}$ obtained over iterations is taken as the approximation. In this case, the steering loss can be redefined as:

$$\mathcal{J}(z_r) = \|z_r - m_{inv}\| \, . \tag{14}$$

Under the guidance of the loss function, the input tensor $z_r$ is gradually adjusted toward the inverted model $m_{inv}$. In each FWI iteration, the steering loss $\mathcal{J}(z_r)$ is computed and updated once. The self-reinforced mechanism acts as an additional prior to complement the intrinsic regularization in the DIP network, allows the input to incorporate legitimate geophysical prior information of the subsurface, and thus effectively reduces artifacts in previous DIP-FWI caused by random and irrelevant network inputs. While the initial input $z_r^0$ is still initialized with random noise, each iteration in SRDIP-FWI updates both the network parameters $\theta$ and the input $z_r$, which can be divided into the following two stages.

1. **Network parameter update:** $z_r^i$ is fixed, while the network parameters $\theta$ are updated by minimizing the data misfit (equation 6):

$$\theta^{i+1} = \theta^i - \eta_\theta \frac{\partial \phi(m^i)}{\partial \theta}, \qquad (15)$$

where $\eta_\theta$ denotes the learning rate for the network parameters. An intermediate inverted velocity model $\tilde{m}^{i+1}$ is generated by:

$$\tilde{m}^{i+1} = m_0 + f_{\theta^{i+1}}(z_r^i). \qquad (16)$$

2. **Network input update:** The network parameters $\theta^{i+1}$ are fixed, and the intermediate inverted model $\tilde{m}^{i+1}$ is used to update input $z_r^i$ by minimizing the steering loss $\mathcal{J}(z_r^i)$:

$$z_r^{i+1} = z_r^i - \eta_z \frac{\partial \mathcal{J}(z_r^i)}{\partial z_r^i} = z_r^i - \eta_z(z_r^i - \tilde{m}^{i+1}), \qquad (17)$$

where $\eta_z$ is the learning rate for the input tensor. The updated input is then used to produce a new model:

$$m^{i+1} = m_0 + f_{\theta^{i+1}}(z_r^{i+1}). \qquad (18)$$

We stress that the initial model $m_0$ remains fixed during the inversion and is not updated. The learned structural information during the inversion process is embedded in the iteratively updated $z_r$.

By updating the network parameters $\theta$ and the input $z_r$ alternately, SRDIP-FWI realize a self-reinforcement mechanism. Figure 1 illustrates the schematic workflow of SRDIP-FWI, and the detailed process is summarized in Algorithm 1. Subsequent numerical experiments demonstrate that this steering algorithm effectively alleviates the ill-posed nature of FWI and significantly improves the accuracy of multiscale velocity model building.

**SYNTHETIC TESTS**

To evaluate the performance of SRDIP-FWI in different subsurface settings, we first perform numerical tests using part of the BP salt model and the Marmousi2 model. The synthetic test using the full BP salt model is presented in Appendix A. For each model, SRDIP-FWI is compared with multiscale FWI (MSFWI), full-band FWI, and existing DIP-FWI using random noise as the network input. No model regularization is applied for FWI or MSFWI for a straightforward comparison with DIP-FWI and SRDIP-FWI. Both FWI and MSFWI use the nonlinear conjugate gradient method for optimization, and the time-frequency decoupling strategy is used in MSFWI to reduce the risk of cycle skipping (Górszczyk et al., 2017). To simplify the comparison and discussion, only the P-wave velocity model is inverted, whereas density is derived from the initial velocity model using the empirical relation (Gardner et al., 1974) and remains fixed. The frequency bands and time windows for MSFWI are listed in Table 1. It should be emphasized that the raw seismic data without any multiscale strategy such as bandpass filtering or time windowing are directed used in DIP-FWI and SRDIP-FWI.

**BP salt model**

We first test SRDIP-FWI using part of the BP model, which contains a complex salt body encompassed by sediments with varied velocities (Billette and Brandsberg-Dahl, 2005). The salt boundary and the underneath velocity variation have long been a challenge for FWI (Gomes et al., 2019). The original model is resampled onto a grid of $nx \times nz = 188 \times 68$ with a grid spacing of 133 m (Figure 2a). The acquisition system with full coverage is deployed on the surface, with source interval being 399 m, and receiver interval being 133 m. A Ricker wavelet with a central frequency of 1.2 Hz is used as the source wavelet, and the total recording time is 27 s. A 1D velocity model with velocity increasing linearly in the vertical direction is used as the initial model (Figure 2b). In this test, 1800 iterations are taken for each method to warrant convergence, and the inversion results are shown in Figures 2c–2f. Understandably, full-band FWI fails to recover the complex salt body and surrounding sediments, yielding ubiquitous artifacts due to convergence to a local minimum (Figure 2c). MSFWI effectively recovers a better velocity model (Figure 2d). However, the salt body, particularly the salt bottom, is poorly reconstructed, and the sediment velocity under the salt body is also not accurately recovered (Figure 2d). For DIP-FWI, the approach is merely able to recover the top salt boundary, failing to recover the remaining structures (Figure 2e). This is because the random noise input in DIP-FWI provides no prior subsurface information, and the output relies entirely on the intrinsic spectral bias of the neural network, which is insufficient to constrain the complex salt bodies and deep sediments weakly illuminated by seismic waves. In contrast, SRDIP-FWI not only accurately recovers the shape and interior velocity of the salt body, but also clearly highlights the velocity contrast across the boundary, which is rather difficult for traditional waveform inversion schemes. In addition, artifacts beneath the salt body are also effectively suppressed by SRDIP-FWI (Figure 2f).

We further quantify the performances of four different FWI approaches in both the data and model domains (Figure 3). In both domains, SRDIP-FWI shows the best performance, and its advantage is particularly pronounced in the model domain (Figure 3b). Table 2 presents three metrics (RMSE, SSIM, and PSNR) in the model domain, and their detailed definitions and computations are provided in Appendix B. The results indicate that SRDIP-FWI achieves the best performance in both structural consistency (SSIM) and inversion accuracy (RMSE, PSNR).

To further evaluate the guidance and impact on convergence of the prior information in SRDIP-FWI, we show the inverted velocity models at different iterations by MSFWI, DIP-FWI and SRDIP-FWI (Figure 4). At the early stage (200th iteration), MSFWI has the best performance among all approaches, which primarily updates the lower-wavenumber components of the velocity model; DIP-FWI roughly captures the top salt boundary at the cost of numerous artifacts; SRDIP-FWI recovers only the top salt boundary, and the velocities of the sediments are minorly updated. At the 400th iteration, MSFWI further enhances the low-wavenumber components in the velocities, but velocity anomalies begin to appear near the 6 km depth; DIP-FWI further strengthens the salt boundaries, while introducing additional smearing and erroneous contrasts simultaneously; SRDIP-FWI successfully delineates the top salt boundary without introducing noticeable artifacts in the low-wavenumber structures. As inversion continues to the 800th iteration, MSFWI gradually refines the salt boundaries, but also introduces artifacts in the shallow regions; DIP-FWI, however, fails to recover the salt body; in comparison, SRDIP-FWI successfully recovers the entire salt body without ringing along the boundaries, and the sediment structures below the salt body also appears. At the 1800th iteration, while MSFWI can recover the primary structures of the salt body, the interior velocity is not inverted accurately, and artifacts are pervasive in the model; DIP-FWI only recovers the top salt boundary, failing to recover salt body and surrounding sediment velocities; in comparison, SRDIP-FWI not only delineate a distinct salt boundary and obtain a quite uniform interior velocity, but also successfully recover the velocities and structures of the surrounding sediments. Overall, by applying the steering algorithm to update the network input, SRDIP-FWI can fully utilize the inverted model to alternately update the input prior and network parameters of the reparameterizing network at each iteration. The feedback mechanism allows the network to continuously incorporate and reinforce structural information during inversion, thereby substantially improving the representational capability of the network for complex structures. It should be stressed that, unlike MSFWI (Table 1), SRDIP-FWI does not require manual selections of frequency bands or time windows. The multiscale model reconstruction in SRDIP-FWI is automatically realized through the spectral bias in the reparametarizing neural network, which is further discussed in the following discussion.

**Marmousi model**

We also benchmark SRDIP-FWI on the Marmousi model (Martin et al., 2006), which contains significant velocity variations and complex fault structures. The Marmousi model is resampled onto a grid of $nx \times nz = 200 \times 71$, with a grid spacing of 40 m (Figure 5a). A Ricker wavelet with a dominant frequency of 5 Hz is used as the source wavelet, with a total recording time of 4.8 s. In total, 40 sources are evenly distributed at 200 m intervals, and 200 receivers are deployed with 40 m intervals along the surface. The initial model (Figure 5b) is obtained using the wave-equation traveltime inversion (Luo et al., 1991) starting from a 1D velocity model, and subsequent inversions with FWI, MSFWI, DIP-FWI and SRDIP-FWI are each performed for 1300 iterations. The results from conventional FWI and MSFWI exhibit pronounced low-velocity anomalies at the shallow margins (dashed circles in Figures 5c and 5d); also, at the depth around 2.5 km (dashed boxes in Figures 5c and 5d), the pinchout bends downward with overly estimated velocities. While DIP-FWI recovers the high-velocity pinchout more accurately, in the depth range of 1.5–3.0 km (dashed ellipses in Figures 5e and 5f), the velocities for the interbedding layers are not accurate, and some scattered artifacts are present. In contrast, SRDIP-FWI achieves high-fidelity reconstruction of both the shallow and deep velocity structures, with significantly less artifacts than the other three FWI approaches (Figure 5f). Further comparisons of the inversion metrics are provided in Appendix C.

**APPLICATION TO THE LAND DATASET ACQUIRED IN INNER**

**MONGOLIA, CHINA**

We further apply SRDIP-FWI to a 2D land seismic dataset acquired in the southwestern section of the Tiancao Depression in Inner Mongolia, China. The result obtained by SRDIP-FWI is compared to that obtained by the traditional MSFWI for the same dataset (Zheng et al., 2024; Liu et al., 2025). The roll-along acquisition on a relative flat terrain consists of 109 shots at a 250-m interval and 409 receivers at a 25-m interval. A preprocessing workflow is conducted to enhance SNR, which includes bad-trace removal, static correction, spherical divergence compensation, and suppressions of surface waves and ambient noise. After preprocessing, early-arrival waveforms are extracted for subsequent waveform inversion. Figures 6a and 6b show the comparison of a shot gather before and after the above preprocessing workflow.

Starting from a 1D layered model, wave-equation traveltime inversion (Luo et al., 1991) is first applied to build an initial velocity model for subsequent FWI (Figure 7a). The maximum time window and frequency ranges used in inversion are determined according to SNR and frequency characteristics of the observed data. The minimum frequency is set to 4 Hz since the signal below is practically absent in the recorded data, and the maximum frequency is set to 16 Hz due to low SNR above. For MSFWI, the inversion is divided into six stages with 2000 iterations in total (Figure 7b), and details for the parameters are listed in Table 3. Both DIP-FWI and SRDIP-FWI directly use the entire frequency band (4–16 Hz) and full time window of MSFWI, with 2000 iterations performed as well (Figure 7c and 7d).

While MSFWI recovers the primary geological structures well, significant smearing and crossovers are also introduced in the inverted model, such as regions indicated by the dashed boxes in Figure 7b. Due to absence of sufficient prior information in the network input, the inversion result by DIP-FWI contains pronounced noise (Figure 7c), indicating that the network struggles to constrain complex geological structures. In contrast, the velocity model yielded by SRDIP-FWI contains more contiguous structures with distinct layer boundaries (Figure 7d). As highlighted by the dashed box in Figure 7d, the artifacts and noise in the inverted model are much less pronounced. Shot gathers at 11.2 km and 32.0 km along the profile (red stars in Figure 7d) are selected for evaluating the similarity between observed and synthetic waveforms. It is found that SRDIP-FWI (Figures 8b and 8d) obtains superior waveform matching in both phase and amplitude compared to MSFWI (Figures 8a and 8c). Figure 9 shows the distributions of data residuals corresponding to the initial model and the final models obtained by MSFWI and SRDIP-FWI. Compared with MSFWI, SRDIP-FWI effectively improves the quality of data matching, especially for those traces with relatively higher similarities (global correlation norm<-0.5).

We further evaluate the accuracy of the inverted models by MSFWI and SRDIP-FWI with prestack depth migration (PreSDM) (Fig. 10). The migration image based on the velocity model by the SRDIP-FWI exhibit more distinct and continuous reflections in several key zones (red arrows in Figure 10). Figures 10c and 10d show the overlaid PreSDM images and corresponding velocity models to better compare the structural consistency. Figure 11 shows the offset-domain common image gathers (ODCIGs) extracted from offsets ranging from 9 km to 17 km. It is obvious that the migrated events in ODCIGs based on the SRDIP-FWI model are more flattened than those with the MSFWI model.

To assess the spatial resolution of SRDIP-FWI, we further conduct a checkerboarding resolution test. Velocity perturbations with 500 m × 500 m in dimension and amplitude of $\pm 500$ m/s (Figure 12e) are added to the model derived by wave-equation traveltime inversion (Figures 7a and 12a) to construct the checkerboard model (Figure 12b) (Spakman and Nolet, 1988; Zheng et al., 2024). "Observed" data are then generated by forward modeling using this checkerboard model for subsequent inversions. Both MSFWI and SRDIP-FWI follow the same workflow and parameter settings as in the inversion with the field data. Figure 12c shows the checkerboard test result obtained by MSFWI, Figure 12d shows the result by SRDIP-FWI, and Figures 12f and 12g show the recovered perturbations against the background model. In general, both FWI approaches recover the checkerboard pattern well in the central region shallower than 1.5 km. However, at the marginal areas (0–5 km and 35–40 km), SRDIP-FWI (Figures 12d and 12g) yields significantly superior recovery in amplitudes and boundaries of the checkerboards. In the region deeper than 1.5 km, the performance of MSFWI degrades noticeably, exhibiting large-scale, low-velocity anomalies and severe structural smearing (Figures 12c and 12f), whereas SRDIP-FWI accurately recovers the checkerboard structures in this deep region (Figures 12c and 12f). As manifested in the checkerboard test, the advantage of SRDIP-FWI can be attributed to the intrinsic network structure, which effectively applies an implicit preconditioning to the waveform inversion. The detailed mechanism is discussed in the following discussion.

## DISCUSSION

### Spectral bias in SRDIP-FWI

The superior performance of SRDIP-FWI over traditional FWI approaches can be attributed to the spectral bias of the reparameterizing neural network, which tends to learn the low-wavenumber components (the smooth, slowly varying structures) of the model first, and learns the high-wavenumber components (rapid varying structures, interfaces) more slowly (Shi et al., 2022) during the inversion. For a neural network $f_\theta(z)$ with ReLU activation, the network function can be explicitly represented as a piecewise-linear summation (Rahaman et al., 2018):

$$f_\theta(z) = \sum_\epsilon 1_{P_\epsilon}(z)(W_\epsilon z + b),$$ (19)

where $\epsilon$ is an index for the linear regions $P_\epsilon(z)$ and $1_{P_\epsilon}(z)$ is the indicator function on $P_\epsilon(z)$, $W_\epsilon$ denotes the weight parameters, and $b$ is the bias term. By applying the Fourier transform, this representation can be expressed in the frequency domain as:

$$\tilde{f}_\theta(\mathbf{k}) = \sum_n \frac{C_n(\theta, \mathbf{k}) 1_{H_n^\theta}(\mathbf{k})}{k^{n+1}},$$ (20)

where $\mathbf{k}$ is the wavenumber, $k = \|\mathbf{k}\|$, $C_n(\theta, \mathbf{k})$ represents the amplitude modulation term, and $1_{H_n^\theta}(\mathbf{k})$ is also an indicator function that captures the directional dependence of spectral decay in the Fourier domain. Since $\tilde{f}_\theta(\mathbf{k})$ contains the factor $k^{-(n+1)}$, we have:

$$\partial \tilde{f}_\theta(\mathbf{k})/\partial \theta = O(k^{-n}).$$ (21)

This equation shows that the gradient of the network output with respect to its parameters $\partial \tilde{f}_\theta(\mathbf{k})/\partial \theta$ is inversely proportional to the frequency, which means higher-wavenumber components are updated more slowly, and lower-wavenumber components are learned first. Thus, the spectral bias guides DIP-FWI and SRDIP-FWI to gradually fill in details with iterations. We further use the frequency band consistency (FBC) metric (Shi et al., 2022), which measures the changes of different wavenumber components in the velocity model, to quantify the spectral bias behavior in SRDIP-FWI. The inverted velocity model is transformed to the wavenumber domain via the 2D Fourier transform:

$$M_i(k_x, k_z) = \iint_{-\infty}^{+\infty} m_i(x, z) e^{-i(k_x x + k_z z)} \, \mathrm{d}x \mathrm{d}z, \tag{22}$$

where $m_i(x, z)$ is the inverted model at the i-th iteration, $M_i(k_x, k_z)$ is its spectral representation in the wavenumber domain, $x$ and $z$ are the spatial coordinates, $k_x$ and $k_z$ are the corresponding wavenumber coordinates. $M_i(k_x, k_z)$ is further divided into $N$ non-overlapped concentric wavenumber bands $M_i^B(k_x, k_z)$:

$$M_i^B(k_x, k_z) = W^B(k_x, k_z) M_i(k_x, k_z), \tag{23}$$

where $W^B(k_x, k_z)$ is the bandpass filter in the wavenumber domain, and $B = 1,2,3 \ldots N$. Figure 13a shows the wavenumber spectrum of the BP salt model (Figure 2a), which is divided into five non-overlapped concentric bands for different wavenumbers (B1 – B5). Assume $M_{true}^B(k_x, k_z)$ is the bandpass-filtered 2D Fourier spectrum of the true velocity model $m_{true}(x, z)$, the FBC for band $B$ is defined as:

$$FBC_B = \frac{\sum_{j=0}^J |M_i^{B,j}(k_x, k_z)|}{\sum_{j=0}^J |M_{true}^{B,j}(k_x, k_z)|} , \tag{24}$$

where $J = n_x \times n_z$, and $n_x$ and $n_z$ are the numbers of grid points in the horizontal and vertical directions. FBC can characterize the convergence behavior of the inverted model at different wavenumbers, and provides a measurement of the spectral learning performance of the neural network.

Figures 13b-e shows the changes of FBC metrics with iterations for the BP salt model (Figure 2) using the four different FWI approaches. The slope of a FBC curve is proportional to the update rate of a corresponding wavenumber component. Without bandpass filtering on seismic data, the update rates of different wavenumber components are almost identical in full-band FWI (Figure 13b). The update rates in MSFWI show discernable differences during the first 300 iterations (Figure 13c), primarily because only low-frequency data are injected in the early iterations, while high-frequency data are gradually included in later iterations. Thus, FBCs of different wavenumber bands gradually converge to the same rate at later iterations. By leveraging the spectral bias in reparameterizing neural networks, DIP-FWI (Figure 13d) exhibits a distinctly faster update rate for the lower-wavenumber components in the early iterations even when the full-band seismic data are used for inversion. However, since random noise is used as the network input, DIP-FWI struggles to recover the structural information in the velocity model, and as iterations proceed, the update rates among different wavenumber bands also begin to converge. In comparison, the lower-wavenumber components are constantly updated more rapidly than the higher-wavenumber components throughout the inversion in SRDIP-FWI (Figure 13d). As a result, the inversion stability is automatically enhanced even without the frequency controls required in traditional MSFWI.

**Advantage of (SR)DIP-FWI compared to conventional FWI**

A key capability of DIP-FWI or SRDIP-FWI lies in exploiting the spectral bias in the neural network to prioritize reconstruction of low-wavenumber components of a velocity model. This feature allows automatic exploitation of different information in the waveforms, without relying on the manually defined frequency bands or time-windows required in conventional FWI. To demonstrate how spectral bias works in matching seismic waveforms, we compare a single trace calculated with the inverted BP salt models at different iterations with SRDIP-FWI and conventional FWI (Figures 2c and 2f). When the entire wave trains are directly used in inversion, the conventional FWI attempts to match all wiggles simultaneously (Figure 14a). However, the inversion gets stuck in a local minimum, and waveforms beyond 14 s are poorly matched. In contrast, though fed with the same unfiltered and un-windowed waveforms as FWI, SRDIP-FWI attempts to match the early arrivals in the first few hundreds of iterations, and then gradually tries to match wiggles arriving later (Figure 14b).

The spectral analysis further reveals how the frequency contents are matched in FWI and SRDIP-FWI with iterations. Since the conventional FWI updates all frequency contents simultaneously, it cannot first utilize the low-frequency information to construct the large-scale background structures to avoid local minima (Figure 14c). In comparison, during the early iterations, the lower frequency contents (<2.0 Hz) are preferentially updated in SRDIP-FWI, while update of higher frequency contents is suppressed by the spectral bias in the neural network (Figure 14d). It is evident that even without bandpass filtering and time windowing, SRDIP-FWI can progressively exploit increasing temporal and frequency contents in the waveforms to reconstruct a velocity model with finer details, thereby inherently stabilizing the inversion process.

**Why (SR)DIP-FWI better reconstruct deep structures**

As evidenced in Figures 4 and 12, (SR)DIP-FWI can better reconstruct the deeper region of the velocity models than the conventional approaches. This improvement arises from the intrinsic mechanism of model reparameterization. In (SR)DIP-FWI, the reparameterized velocity model is represented in a much more compact latent space. During the iterative optimization, the depth-sensitive gradients are converted into depth-insensitive gradients represented by the network parameters, thereby mitigating the discrepancy in depth illumination. That is, the four-level U-Net encodes the input model ($188 \times 68$) into a feature map ($11 \times 4$) that is significantly smaller (Figure 1). Figure 15 shows the evolution of the feature map and the inversion results at different iterations, which clearly demonstrates how the velocity model is compressed and reconstructed through the encoder-decoder architecture. The encoder effectively compresses both the vertical and horizontal dimensions of the inverted velocity model, yielding a more compact spatial representation which promotes deep structure update during inversion.

To further examine how the encoder-decoder process can facilitate velocity recovery, we conduct an ablation test using a single-level U-Net in the reparameterizing DIP network based on the partial BP model (Figures 16a and 16b). The inversion is also performed for 1800 iterations. In this case, the original velocity model ($188 \times 68$) is encoded into a feature map with a spatial dimension of $94 \times 34$ (Figure 16a), the spatial compression of which is significantly less compared to the feature map encoded by the four-level U-Net (Figure 15). Figures 16b shows the inverted velocity models reparameterized by the single-level DIP network, the deep structure of which is recovered much worse compared to that reparameterized by the four-level U-Net (Figure 15f). Also, since the four-level U-Net and the single-level U-Net differ substantially in their numbers of network parameters, we increase the channel of the single-level U-Net such that the total number of its parameters (9,452,985) is comparable to that of the four-level U-Net (9,432,193). Figure 16c shows the bottleneck feature map of the expanded single-level U-Net, and Figure 16d shows the corresponding inverted velocity model. The comparison between Figures 16d and 15f apparently suggests that recovery of deep structures depends primarily on the depth rather than the number of parameters of the DIP network.

## ACKNOWLEDGEMENT

This study is funded by the National Natural Science Foundation of China under grants U2139204, and by the Deep Earth Probe and Mineral Resources Exploration-National Science and Technology Major Project 2025ZD1007502. The original code of SRDIP-FWI, along with synthetic datasets, scripts, and instructions to reproduce the results in this study, is publicly available at GitHub to ensure full transparency and reproducibility of the work.

## CONCLUSION

In this study, we introduce a novel waveform inversion framework named self-reinforced deep image prior-based FWI (SRDIP-FWI). This new approach not only reparameterizes the subsurface velocity model using a U-Net architecture, which embeds implicit regularization through the spectral bias of the network, but also incorporates a steering algorithm that alternately updates both the network parameters and its input during iterative inversion. Through extensive synthetic tests involving models with sharp velocity contrasts and complex structures, we demonstrate that SRDIP-FWI can effectively alleviate the ill-posedness inherent in conventional FWI and existing DIP-FWI. The effectiveness of the proposed method is further validated on a real land seismic dataset from Inner Mongolia, China. With the FBC metric, we also reveal how the spectral bias of the reparameterizing neural network can prioritize update of the lower-wavenumber components to improve stability of the iterative inversion. It should be emphasized that the proposed SRDIP-FWI can automatically perform adaptive multiscale inversion without explicit bandpass filtering and time windowing of the seismic data, thereby greatly simplifying the current FWI workflow. We also reveal why SRDIP-FWI can better reconstruct regions with weak illuminations compared to conventional approaches by showing how the traces are progressively matched and how the velocity model is mapped into a more compact latent space. We hope that SRDIP-FWI will serve as a powerful and practical tool for high-fidelity reconstruction of complex subsurface structures while significantly simplifying the current inversion workflows.

## APPENDIX A

**Synthetic test with the Full BP Model**

We further evaluate the performances of different methods using the full BP

model. In this test, the full BP model is resampled to $n_x \times n_z$=401×91 with a grid spacing of 62.5 m (Figure A-1a). A Ricker wavelet with a central frequency of 2.3 Hz is used as the source wavelet, with a time sampling interval of 4 ms and a total recording duration of 20 s. In total, 81 shots are evenly distributed along the surface with a spacing of 180 m, and 151 receivers are evenly placed on the surface with a spacing of 312.5 m. A 1D velocity model is used as the initial model (Figure A-1b). A total of 3000 iterations are conducted for each waveform inversion approach, and the results are shown in Figures A-1c to A-1f. The conventional FWI easily gets trapped in local minima due to an inaccurate initial model, poorly recovering the salt body with ubiquitous artifacts (Figure A-1c). MSFWI recovers an improved velocity model, but the salt bottom remains obscure and noticeable artifacts also appear below the salt (Figure A-1d). The existing DIP-FWI partially recovers the top salt boundary and the interior structure at shallower depths, while the deeper structure is poorly recovered with apparent noise (Figure A-1e). In contrast, SRDIP-FWI recovers the entire salt body and the sediment velocities successfully with minor artifacts only at the deepest section of the model. Particularly, the salt boundary is quite sharp (Figure A-1f).

Figure A-2a shows the global correlation norm with iterations for different FWI approaches. It is evident that SRDIP-FWI has better performance over the other three approaches. The difference among different FWI approaches becomes more pronounced in the model domain (Figure A-2b): the models derived by conventional FWI and MSFWI have substantially higher errors, while DIP-FWI converges relatively rapidly in the early iterations, but stagnates later. In comparison, the error of the velocity

model derived by SRDIP-FWI continues to reduce even in later iterations, yielding a distinctly better model than the other approaches. The SSIM and PSNR metrics listed in Table A-1 also indicate that SRDIP-FWI obtains higher structural similarity and reconstruction quality of the velocity model compared to the other approaches.

## APPENDIX B

### Definitions of RMSE, PSNR, and SSIM

We quantitatively evaluate the accuracy of the inverted models using three metrics: the root-mean-square error (RMSE), peak signal-to-noise ratio (PSNR), and structural similarity index (SSIM). RMSE is defined as:

$$RMSE(m_{inv}, m_{true}) = \frac{1}{\sqrt{n_x n_z}} \|m_{inv} - m_{true}\|_2 , \qquad (B-1)$$

where $m_{inv}$ is the inverted model, and $m_{true}$ is the true model.

PSNR quantifies the similarity between the inverted and true models, with higher values indicating greater similarity. PSNR is defined as:

$$PSNR = 20 \log_{10} \left( \frac{MAX(m_{true})}{\sqrt{RMSE(m_{inv}, m_{true})}} \right), \qquad (B-2)$$

where $MAX(m_{true})$ is the maximum value of the true model.

SSIM also quantifies the structural similarity between the true and inverted models, with larger values indicating higher similarity. SSIM is defined as:

$$SSIM = \frac{(2\mu_{true}\mu_{inv} + c_1)(2\sigma_{true,inv} + c_2)}{(\mu_{true}^2 + \mu_{inv}^2 + c_1)(\sigma_{true}^2 + \sigma_{inv}^2 + c_2)} , \qquad (B-3)$$

where $\mu_{true}$ and $\mu_{inv}$ are the mean values of the true and inverted models, respectively, $\sigma_{true}^2$ and $\sigma_{inv}^2$ are the variances of the true and inverted models, respectively, $\sigma_{true,inv}$ is the covariance, and $c_1 = (0.01 \cdot (v_{max} - v_{min}))^2$ and $c_2 = (0.03 \cdot (v_{max} - v_{min}))^2$ are constants, with $v_{max}$ and $v_{min}$ being the maximum and minimum velocities in the true model, respectively.

## APPENDIX C

**Supplementary results for the Marmousi Model Test**

Figure C-1 shows the global correlation norm for different FWI approaches on the Marmousi model (Figure 5). While all methods produce similar data misfits (Figure C-1a), SRDIP-FWI obtains a significantly better velocity model (Figure C-1b) compared to conventional FWI and MSFWI. The SSIM and PSNR metrics listed in Table C-1 further demonstrate that SRDIP-FWI achieves higher structural similarity and reconstruction quality of the velocity model compared to the other approaches.

## REFERENCES

Adamczyk, A., M. Malinowski, and A. Górszczyk, 2015, Full-waveform inversion of conventional Vibroseis data recorded along a regional profile from southeast Poland: Geophysical Journal International, 203, no. 1, 351–365, doi:10.1093/gji/ggv276

Barnier, G., E. Biondi, and R. Clapp, 2019, Waveform inversion by model reduction using spline interpolation, SEG, Expanded Abstracts, 1400–1404, doi: 10.1190/segam2019-3216866.1.

Billette, F. J., and S. Brandsberg-Dahl, 2005, The 2004 BP Velocity Benchmark: In 67th EAGE Conference & Exhibition. European Association of Geoscientists & Engineers.

Biondi, B., 1992, Velocity estimation by beam stack: Geophysics, 57, no. 8, 1034-1047, doi: 10.1190/1.1443315

Boonyasiriwat, C., P. Valasek, P. Routh, and X. Zhu, 2009, Application of multiscale waveform tomography for high-resolution velocity estimation in complex geologic environments: Canadian foothills synthetic data example: The Leading Edge, 28, no. 4, 454-456, doi: 10.1190/1.3112764

Bozdağ, E., J. Trampert, and J. Tromp, 2011, Misfit functions for full waveform inversion based on instantaneous phase and envelope measurements: Geophysical Journal International, 185, no. 2, 845–870, doi: 10.1111/1365-246X. 2011. 04970.x

Bunks, C., F. M. Saleck, S. Zaleski, and G. Chavent, 1995, Multiscale seismic waveform inversion: Geophysics, 60, 1457–1473, doi: 10.1190/1 .1443880.

Cheng, S., Y. Wang., Q., Zhang, R. Harsuko, and T. Alkhalifah, 2025, A Self-Supervised Learning Framework for Seismic Low-Frequency Extrapolation : Journal of Geophysical Research: Machine Learning and Computation, 1, no. 3, doi: https://doi.org/10.1029/2024JH000157

Choi, Y., and T. Alkhalifah, 2012, Application of multi-source waveform inversion to marine streamer data using the global correlation norm: Geophysical Prospecting, 60, no. 4, 748-758, doi: 10.1111/1365-2478.2012.01079.x.

Dantas, R. S., W. E. Medeiros, and J. Costa, 2019, A multiscale approach to full-waveform inversion using a sequence of time-domain misfit functions, 84, no. 4, 539–551, doi: 10.1190/geo2018-0291.1

Dhara, A., and M. K. Sen, 2023, Physics-guided deep autoencoder to overcome the need for a starting model in full-waveform inversion: The Leading Edge, 41, no. 6, 375–381, doi: 10.1190/tle41060375.1

Gardner, G. H. F., L. W. Gardner, and A. R. Gregory, 1974, Formation velocity and density—the diagnostic basics for stratigraphic traps: Geophysics, 39, no. 9, 770–780, doi: 10.1190/1.1440465

Gomes, A., Peterson, J., Bitlis, S., Fan, C., & Buehring, R. (2019). Assisting salt model building with reflection full-waveform inversion. Interpretation, 7(2), SB43–SB52. https://doi.org/10.1190/INT-2018-0155.1

Górszczyk, A., S. Operto, M. Malinowski,  2017, Toward a robust workflow for deep crustal imaging by FWI of OBS data: The eastern Nankai Trough revisited: Journal of Geophysical Research: Solid Earth, 122, no. 6, 4601-4630, doi: 10.1002/2016JB013891.

He, Q., and Y. Wang, 2021, Reparameterized full-waveform inversion using deep neural networks: Geophysics, 86, no. 1, 1–13, doi: 10.1190/geo2019-0382.1

LeCun, Y., Y. Bengio, and G. Hinton, 2015, Deep learning: Nature, 521(7553), 436–444, doi: 10.1038/nature14539.

Lemaistre, L., J. Brunelliere, F. Studer, and C. Rivera, 2018, FWI on land seismic datasets with topography variations: Do we still need to pick first arrivals?: SEG Expanded Abstracts, 1078-1082, doi: 10.1190/segam2018-2995924.1

Li, H., J. Li, B. Liu and X. Huang, 2021，Application of full-waveform tomography on deep seismic profiling data set for tectonic fault characterization, SEG, Expanded Abstracts, 657-661, doi: 10.1190/segam2018-2995924.1

Li, H., J. Li, S. Luo, T. S. Bem, H. Yao, and X. Huang, 2023, Continent-continent collision between the South and North China plates revealed by seismic refraction and reflection at the southern segment of the Tanlu Fault Zone: Journal of Geophysical Research: Solid Earth, 128, no. 1, e2022JB025748, doi: 10.1029/2022JB025748.

Li, H., R. Clapp, 2024, Time-lapse full-waveform inversion by model order reduction using radial basis function, SEG, Expanded Abstracts, 807-811, doi: 10.1190/image2024-4094488.1

Liu, B., P. Jiang, Q. Wang, Y. Ren, S. Yang, and A. G. Cohn, 2023, Physics-driven self-supervised learning system for seismic velocity inversion: Geophysics, 88, no. 2, 145–161, doi: 10.1190/geo2021-0302.1

Liu, F., H. Li, G. Zou, and J. Li, 2025, Automatic differentiation-based full waveform inversion with flexible workflows: Journal of Geophysical Research: Machine Learning and Computation, 2, no. 1, doi: 10.1029/2024JH000542

Liu, F., Y. Li, R. Su, J. Huang, and L. Bei, Deep Reparameterization for Full Waveform Inversion: Architecture Benchmarking, Robust Inversion, and Multiphysics Extension: arXiv preprint, doi: 10.48550/arXiv.2504.17375

Liu, Y., D. Feng, Y. Xiao, G. Huang, L. Cai, and X. Tai, 2024, Full-Waveform Inversion of Multifrequency GPR Data Using a Multiscale Approach Based on Deep Learning: IEEE Transactions on Geoscience and Remote Sensing, 62, doi: 10.1109/TGRS.2024.3382331

Lu, C., Y. Wang, X. Zou, J. Zong, and Q. Su, 2024, Elastic full-waveform inversion via physics-informed recurrent neural network: IEEE Transactions on Geoscience and Remote Sensing, 62 , 1–16, doi: 10.1109/TGRS.2024.3450696

Lu, K., Y. Wang, 2025, Seismic full-waveform inversion regularized with a migration image: Geophysics, 90, no. 3, 143-157, doi: 10.1190/GEO2023-0419.1

Luo, Y., and G. T. Schuster, 1991, Wave-equation traveltime inversion: Geophysics, 56, no. 5, 645-653, doi: 10.1190/1.1443081.

Pilario, K. E., Shafiee, M., Cao, Y., Lao, L., and Yang, S.-H., 2020, A review of kernel methods for feature extraction in nonlinear process monitoring: Processes, 8(1), 24, https://doi.org/10.3390/pr801002

Martin, G. S., R. Wiley, and K. J. Marfurt, 2006, Marmousi2: An elastic upgrade for Marmousi: The Leading Edge, 25, 156–166, doi: 10.1190/ 1.2172306.

Masmoudi, N., A. Ratcliffe, O. Bukola, J. Tickle, and X. Chen, 2024, Elastic FWI of multi-component ocean-bottom seismic to update shear-wave velocity models: 85th EAGE Annual Conference & Exhibition, 1–5, doi: 10.3997/2214-4609.202410528.

Métivier, L., A. Allain, R. Brossier, Q. Mérigot, E. Oudet, and J. Virieux, 2016, Optimal transport for seismic full waveform inversion: Communications in Mathematical Sciences, 14(8), 2309–2330, https://doi.org/10.4310/CMS.2016.v14.n8.a9

Plessix, R., 2006, A review of the adjoint-state method for computing the gradient of a functional with geophysical applications: Geophysical Journal International, 167,495–503, doi: 10.1111/1365-246X.2006.02978.x.

Plessix, R., G. Baeten, J. W. d. Maag, and F. t. Kroode, 2012, Full waveform inversion and distance separated simultaneous sweeping: a study with a land seismic data set: *Geophysical Prospecting*, 60, no. 4, 733-747, doi: 10.1111/1365-2478.2011.01036.x.

Pratt, R. G., 1999, Seismic waveform inversion in the frequency domain Part 1: Theory and verification in a physical scale model: Geophysics, 64, 888–901, doi: 10.1190/1.1444597.

Prieux, V., G. Lambaré, S. Operto, and J. Virieux, 2012, Building starting models for full waveform inversion from wide-aperture data by stereotomography: Geophysical Prospecting, 109–137, doi: 10.1111/1365-2478.2012.01099.x.

Rumelhart, D. E., G. E. Hinton, and R. J. Williams, 1986, Learning representations by back-propagating errors: Nature, 323, 533–536, doi: 10.1038/323533a0.

Richardson, A., 2018a, Generative adversarial networks for model order reduction in seismic full-waveform inversion: arXiv e-prints, arXiv:1806.00828.

Richardson, A., 2018b, Seismic full-waveform inversion using deep learning tools and techniques: arXiv e-prints, arXiv:1801.07232.

Schouten, T. L., L. Gebraad, S. Neo, J. P. Gülcher, S. Thrastarson, D. v. Herwaarden, A. Fichtner, 2024, Full-waveform inversion reveals diverse origins of lower mantle positive wave speed anomalies: Scientific reports, 14, 26708(2024), doi: 10.1038/s41598-024-77399-2.

Shi, Z., P. Mettes, S. Maji, and C. G. M. Snoek, 2022, On measuring and controlling the spectral bias of the deep image prior: International Journal of Computer Vision, 130, 885-908, doi: 10.1007/s11263-021-01572-7

Shu, Z., Z. Pan, 2025, SDIP: Self-reinforcement deep image prior framework for image processing: Pattern Recognition, 168, doi: 10.1016/j.patcog.2025.111786.

Singh, B., M. Malinowski, A. Górszczyk, A. Malehmir, S. Buske, Ł. Sito, and P. Marsden, 2022, 3D high-resolution seismic imaging of the iron oxide deposits in Ludvika (Sweden) using full-waveform inversion and reverse time migration: Solid Earth, 13, no. 6, 1065–1085, doi:10.5194/se-13-1065-2022.

Sirgue, L., and R. G. Pratt, 2004, Efficient waveform inversion and imaging: A strategy for selecting temporal frequencies: Geophysics, 69, 231–248, doi: 10.1190/1.1649391.

Sirgue, L., O. I. Barkved, J. P. V. Gestel, O. J. Askim, and J. H. Kommedal, 2009, 2D waveform inversion on Valhall wide-azimuth OBS: 71st Conference & Technical Exhibition, EAGE, ExtendedAbstracts, U038.

Song, C., T. Alkhalifah, and U. B. Waheed, 2020, Solving the frequency-domain acoustic VTI wave equation using physics-informed neural networks: Geophysical Journal International, 225, no. 2, 846–859, doi: 10.1093/gji/ggab010

Strauss, R. R., R. Ramanujan, A. Becker, and T. C. Peck, 2020, A steering algorithm for redirected walking using reinforcement learning: IEEE Transactions on Visualization and Computer Graphics, 26, no. 5, 1955–1963, doi:10.1109/TVCG.2020.2968543.

Spakman, W.; Nolet, G. Imaging algorithms, accuracy and resolution in delay time tomography. In Mathematical Geophysics; Springer: Berlin/Heidelberg, Germany, 1988; pp. 155–187.

Sun, J., K. A. Innanen, and C. Huang, 2021, Physics-guided deep learning for seismic inversion with hybrid training and uncertainty analysis: Geophysics, 86, no. 3, 303–317, doi: 10.1190/geo2020-0312.1

Sun, J., K. Innanen, T. Zhang, and D. Trad, 2023, Implicit seismic full waveform inversion with deep neural representation: Journal of Geophysical Research: Solid Earth, 128, no. 3, e2022JB025964, doi: 10.1029/2022jb025964

Tarantola, A., 1984, Inversion of seismic reflection data in the acoustic approximation: Geophysics, 49, 1259–1266, doi: 10.1190/1.1441754.

Tarantola, A., 2005, Inverse Problem Theory and Methods for Model Parameter Estimation: Society for Industrial and Applied Mathematics (SIAM).

Tariq, S., and B. K. P. Horn, 2022, Direct domain adaptation through reciprocal linear transformations: Frontiers in Artificial Intelligence, 5, 927676, https://doi.org/10.3389/frai.2022.927676.

Tromp, J., 2019, Seismic wavefield imaging of Earth's interior across scales: Nature Reviews Earth & Environment, 1, no. 1, 40–53, doi: 10.1038/s43017-019-0003-8.

Ulyanov, D.,A. Vedaldi, and V. Lempitsky, 2018, Deep image prior: In Proceedings of the ieee conference on computer vision and pattern recognition, 9446–9454, doi: 10.1109/CVPR.2018.00984

Wang, D., C. Chen, D. Zhuang, J. Mei, and P. Wang, 2021, Land FWI: Challenges and possibilities: 82nd EAGE Annual Conference & Exhibition, 1–5, doi: 10.3997/2214-4609.202011175.Woodward, M. J., D. Nichols, O. Zdraveva, P. Whitfield, and T. Johns, 2008, A decade of tomography: Geophysics, 73, no. 5, VE5–VE11, doi: 10.1190/1.2969907.

Wu, RS., J. Luo, and B. Wu, 2014, Seismic envelope inversion and modulation signal model: Geophysics, 79, no. 3, 13-24, doi: 10.1190/geo2013-0294.1.

Wu, Y., and G. A. McMechan, 2019, Parametric convolutional neural network-domain full-waveform inversion: Geophysics, 84, no. 6, R881R896, doi: 10.1190/geo2018-0224.1.

Wu, Y., J. Ma, 2025, How Does Neural Network Reparametrization Improve Geophysical Inversion? : Journal of Geophysical Research: Machine Learning and Computation, 2, no. 2, doi: 10.1029/2025JH000621

Wu, Y., Y. Lin, and Z. Zhou, 2018, InversionNet: Accurate and efficient seismic waveform inversion with convolutional neural networks: 88th Annual International Meeting, SEG, Expanded Abstracts, 2096–2100, doi: 10.1190/segam2018-2998603.1.

Yang, F., and J. Ma, 2019, Deep-learning inversion: Anext-generation seismic velocity model building method: Geophysics, 84, no. 4, 583–599, doi: 10.1190/geo2018-0249.1

Zelt, C., and R. B. Smith, 1992, Seismic traveltime inversion for 2-D crustal velocity structure: Geophysical Journal International 108, 16–34, doi: 10.1111/1365-

246X.1992.tb00836.x

Zerafa, C., Galea, P., and Sebu, C., 2025, Synergizing Deep Learning and Full-Waveform Inversion: Bridging Data-Driven and Theory-Guided Approaches for Enhanced Seismic Imaging: arXiv preprint arXiv:2502.17585.

Zhang, Z., and Y. Lin, 2020, Data-driven seismic waveform inversion: A study on the robustness and generalization: IEEE Transactions on Geoscience and Remote Sensing, 58, no. 10, 6900–6913, doi: 10.1109/TGRS.2020.2971635

Zhao, D., F. Zhao, Y. J., Gan, 2020, Reference-Driven Compressed Sensing MR Image Reconstruction Using Deep Convolutional Neural Networks without Pre-Training: Sensors, 20(1), 308, doi: 10.3390/s20010308

Zhao, P., J. Fang, C. Jie, J. Zhang, E. Wang, and S. Zhang, 2025, Multiscale Deep Learning Reparameterized Full Waveform Inversion With the Adjoint Method: IEEE Transactions on Geoscience and Remote Sensing, 63, doi: 10.1109/TGRS.2025.3553053

Zheng, X., J. Li, Q. Xiong, J. Xie, H. Li, J. Tan, Z. Liu, and T. Duan, 2024, Application of active-source full waveform inversion in multi-resource exploration in Juyanhai depression of inner Mongolia: Chinese Journal of Geophysics, 67(8), 3120–3135, doi: 10.6038/cjg2024R0753.

Zhu, Wu., K. Xu, E. Darve, B. Biondi, and G. C. Beroza, 2022, Integrating deep neural networks with full-waveform inversion: Reparameterization, regularization, and uncertainty quantification: Geophysics, 87, no. 1, 93–109, doi: 10.1190/geo2020-0933.1

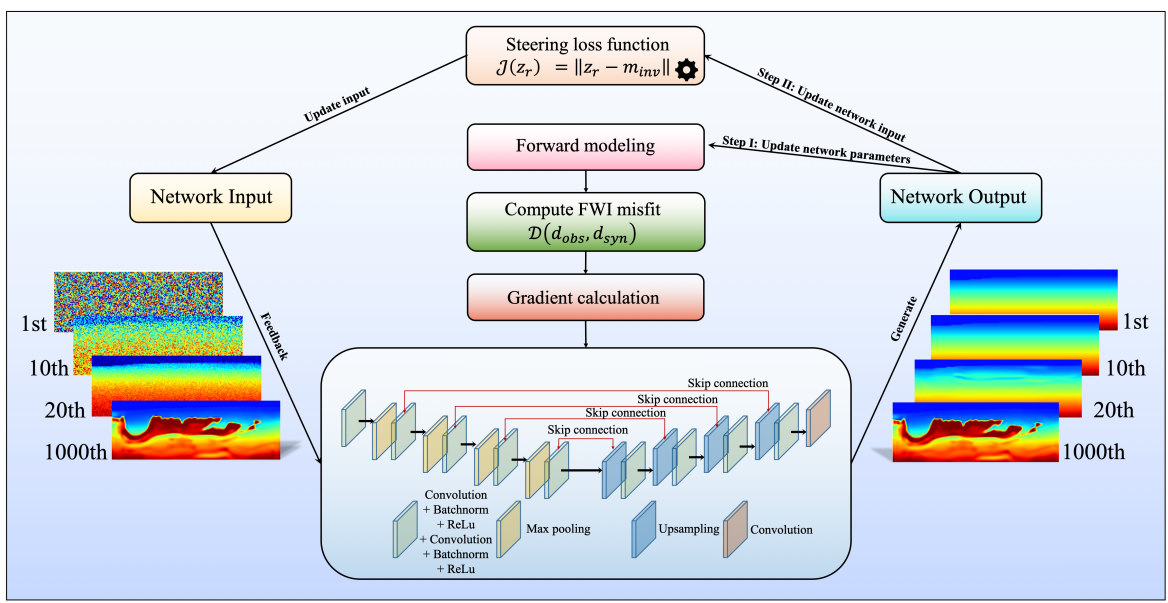

**Figure 1.** Schematic diagram for the self-reinforced deep image prior full waveform

inversion (SRDIP-FWI). A four-level U-Net is used at the reparameterizing DIP

network. For an input tensor of $(n_x, n_z)$, the encoder downsamples the feature map by a factor of 2 at each level, while the decoder upsamples the map by a factor of 2 at each level, constituting a four-level U-Net architecture.

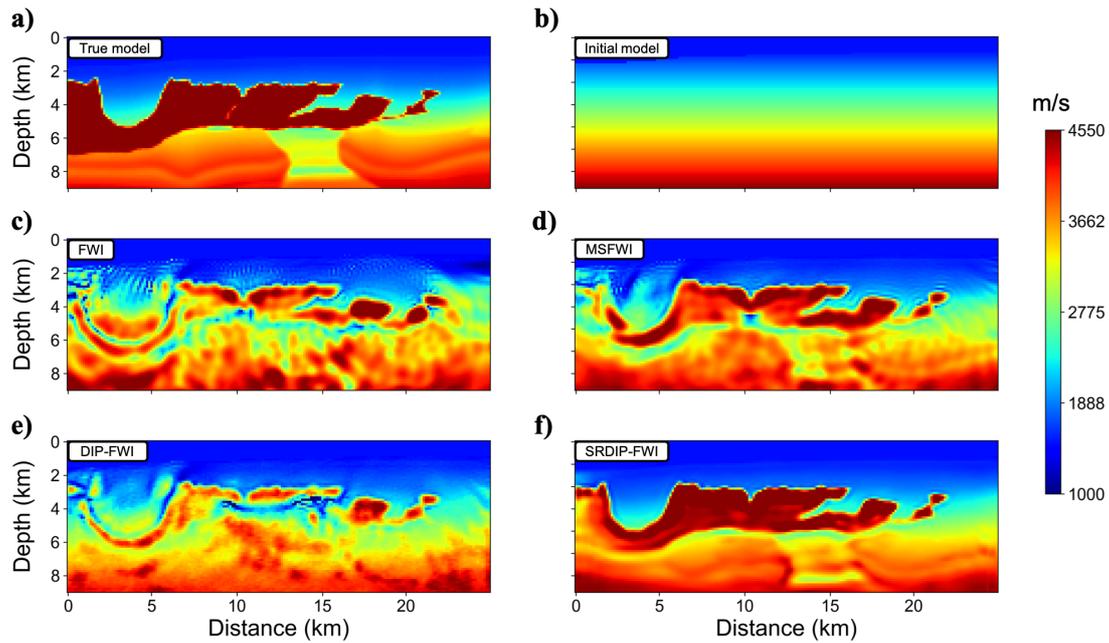

**Figure 2.** Inversion results for part of the BP model. (a) The true model to be recovered, which is cut from the left section of the original BP model; (b) the initial 1D model with velocity increasing linearly in the vertical direction; inversion results using (c) FWI, (d) MSFWI, (e) DIP-FWI, and (f) SRDIP-FWI, respectively.

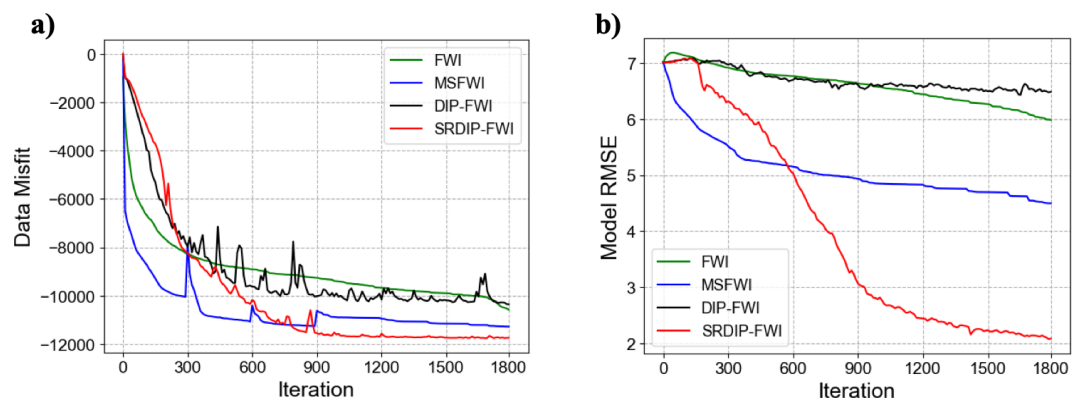

**Figure 3.** Data global correlation norm and model RMSE for the inverted BP model using FWI, MSFWI, DIP-FWI, and SRDIP-FWI. (a) Data global correlation norms, and (b) model RMSEs with iterations for the four different FWI approaches.

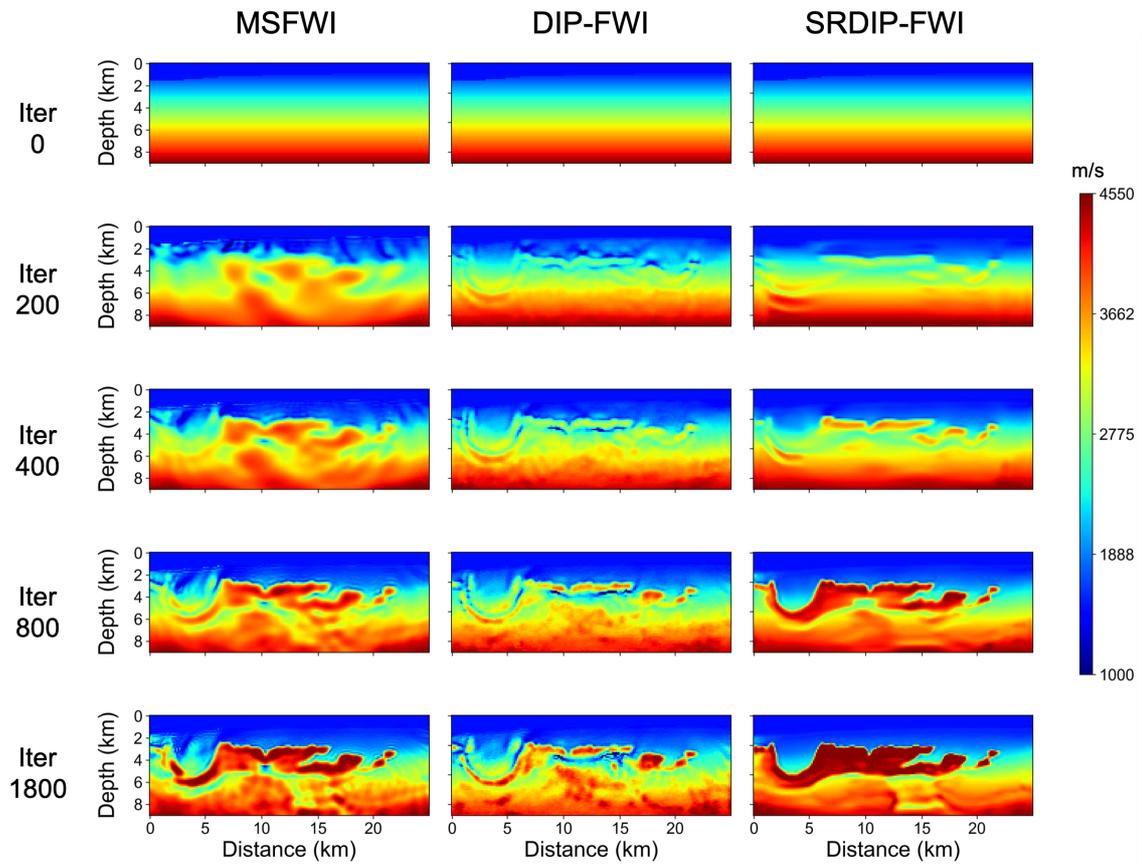

**Figure 4.** Comparison of the inversion results by MSFWI, DIP-FWI, and SRDIP-FWI at the different iterations. The frequency bands and time windows used for MSFWI are listed in Table 1.

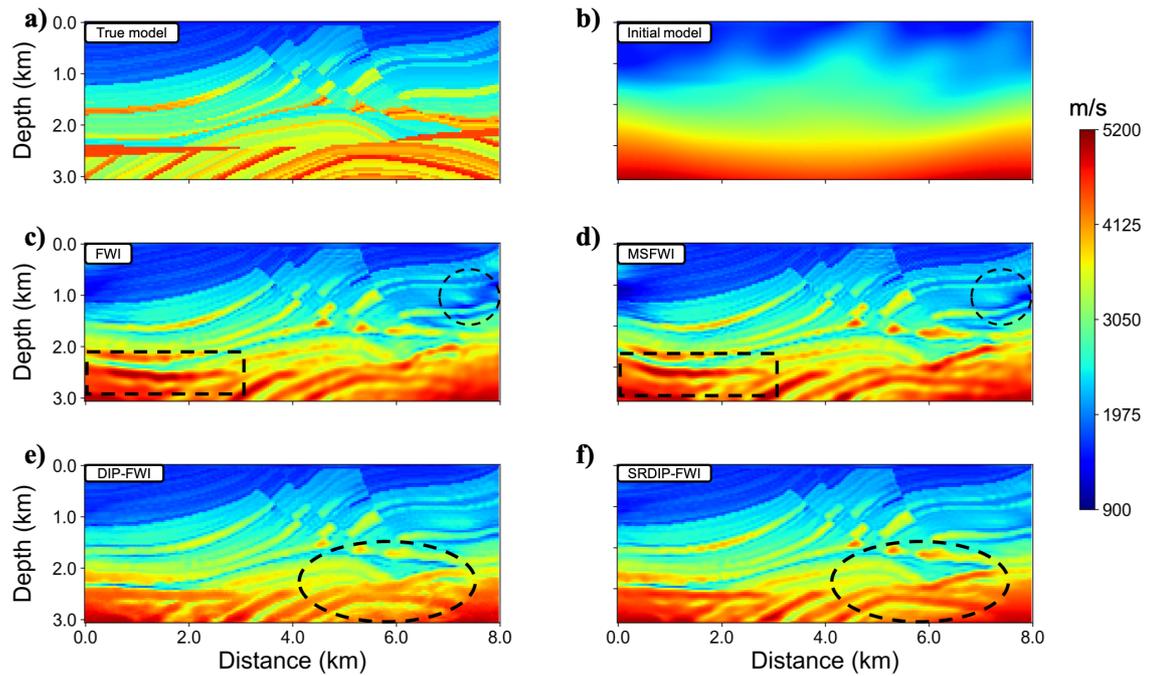

**Figure 5.** Inversion results for the Marmousi model with different FWI approaches.

(a) The true Marmousi model; (b) the initial model derived from wave-equation

traveltime inversion; inversion results using (c) full-band FWI, (d) MSFWI, (e) DIP-

FWI, and (f) SRDIP-FWI, respectively.

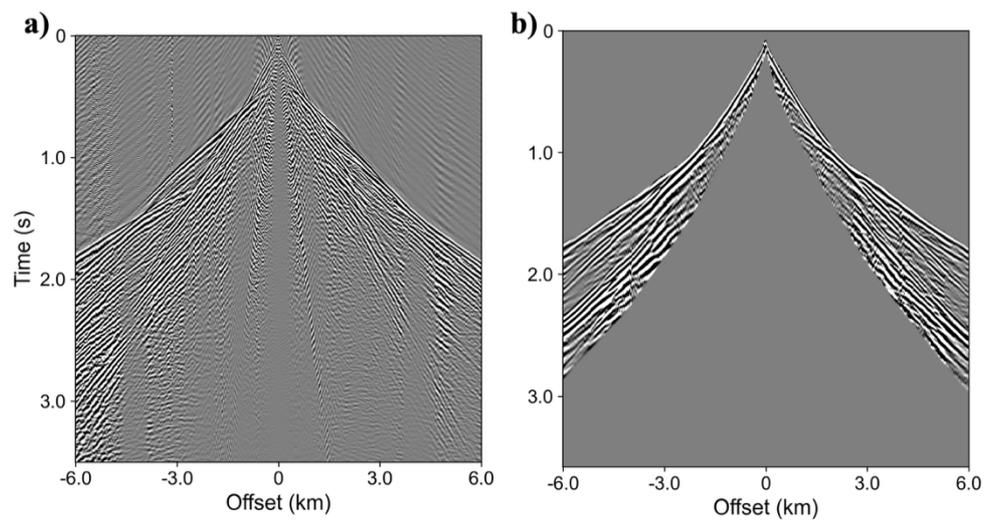

**Figure 6.** Comparison of a shot gather before and after preprocessing. (a) The raw shot gather before preprocessing; (b) the processed shot gather after bad-trace removal, static correction, spherical divergence compensation, and suppressions of surface waves and ambient noise.

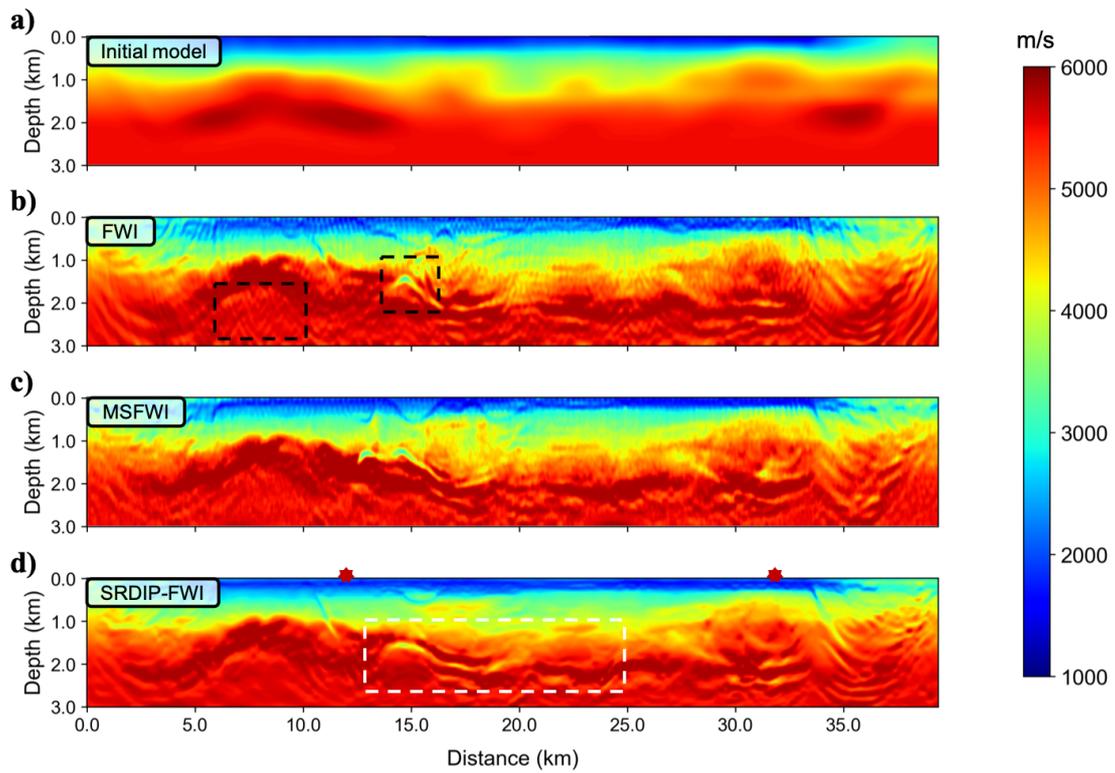

**Figure 7.** Comparison of inversion results with different FWI approaches for the field

dataset acquired in Inner Mongolia, China. (a) The initial velocity model obtained

from wave-equation traveltime inversion; (b) inversion results using MSFWI, (c)

DIP-FWI, (d) SRDIP-FWI, respectively. The two red stars in (d) indicate the locations of the shot gathers shown in Figure 8.

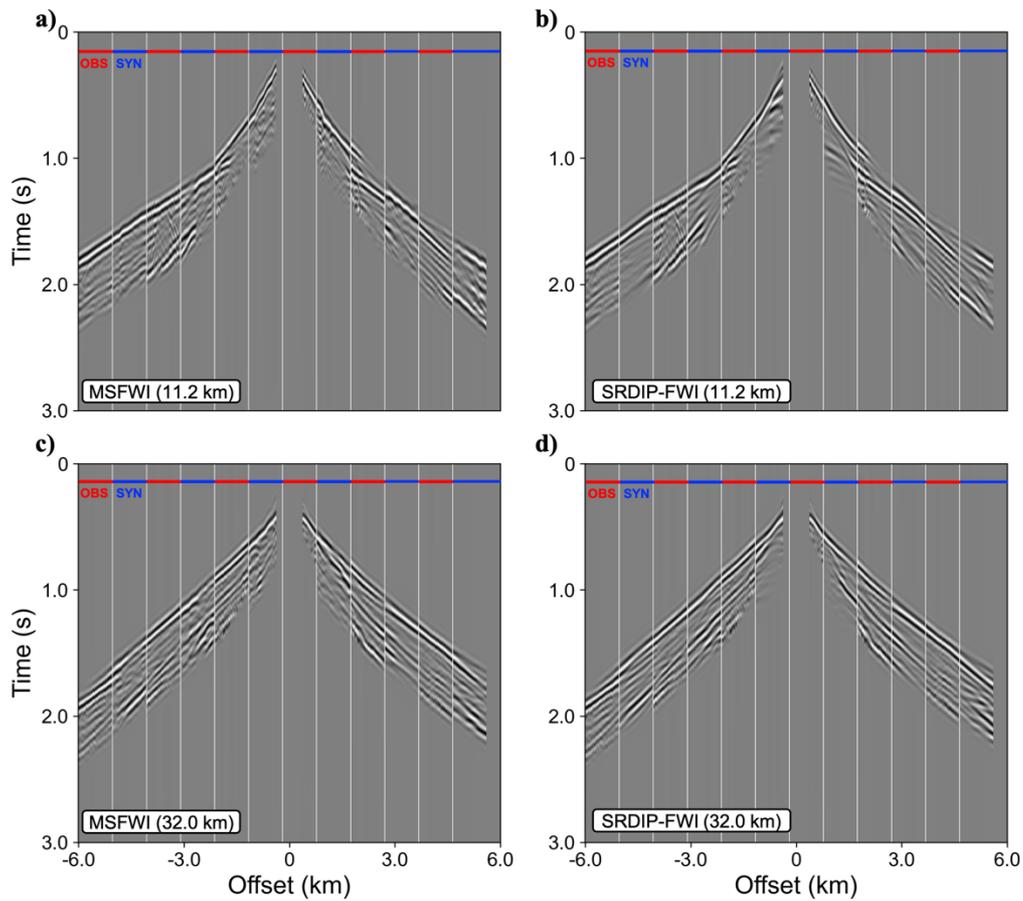

**Figure 8.** Comparison between observed and synthetic waveforms for shot gathers at 11.2 km and 32.0 km along the survey line. The observed (red) and synthetic traces are shown and compared in alternate panels. Comparisons of waveforms at (a) 11.2 km for MSFWI; (b) 11.2 km for SRDIP-FWI; (c) and (d) are similar to (a) and (b), respectively, but for the shot gather at 32.0 km along the survey line.

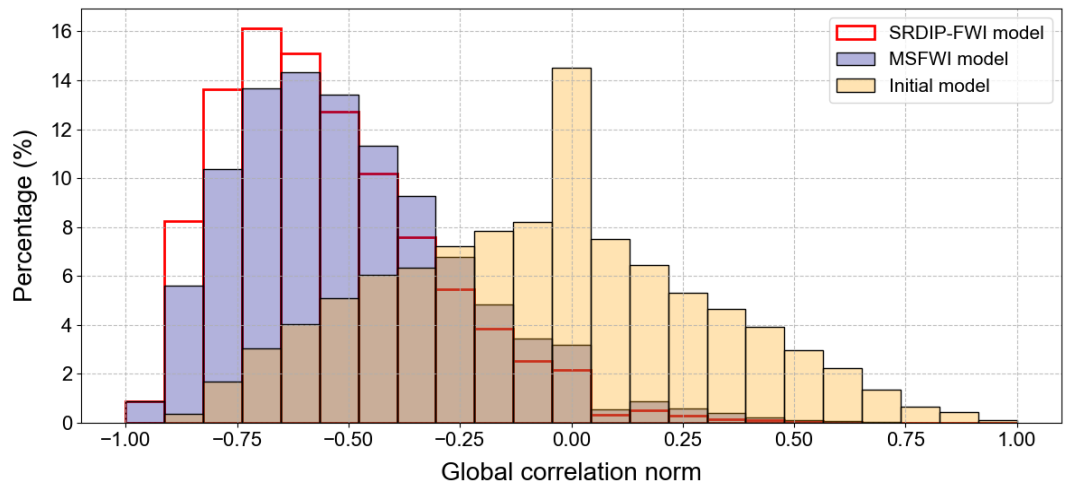

**Figure 9.** Distributions of data residuals corresponding to the initial model and the final models obtained by MSFWI and SRDIP-FWI models. Note the negative global correlation norm is used as the residual metric, and values closer to -1 indicate higher waveform similarity.

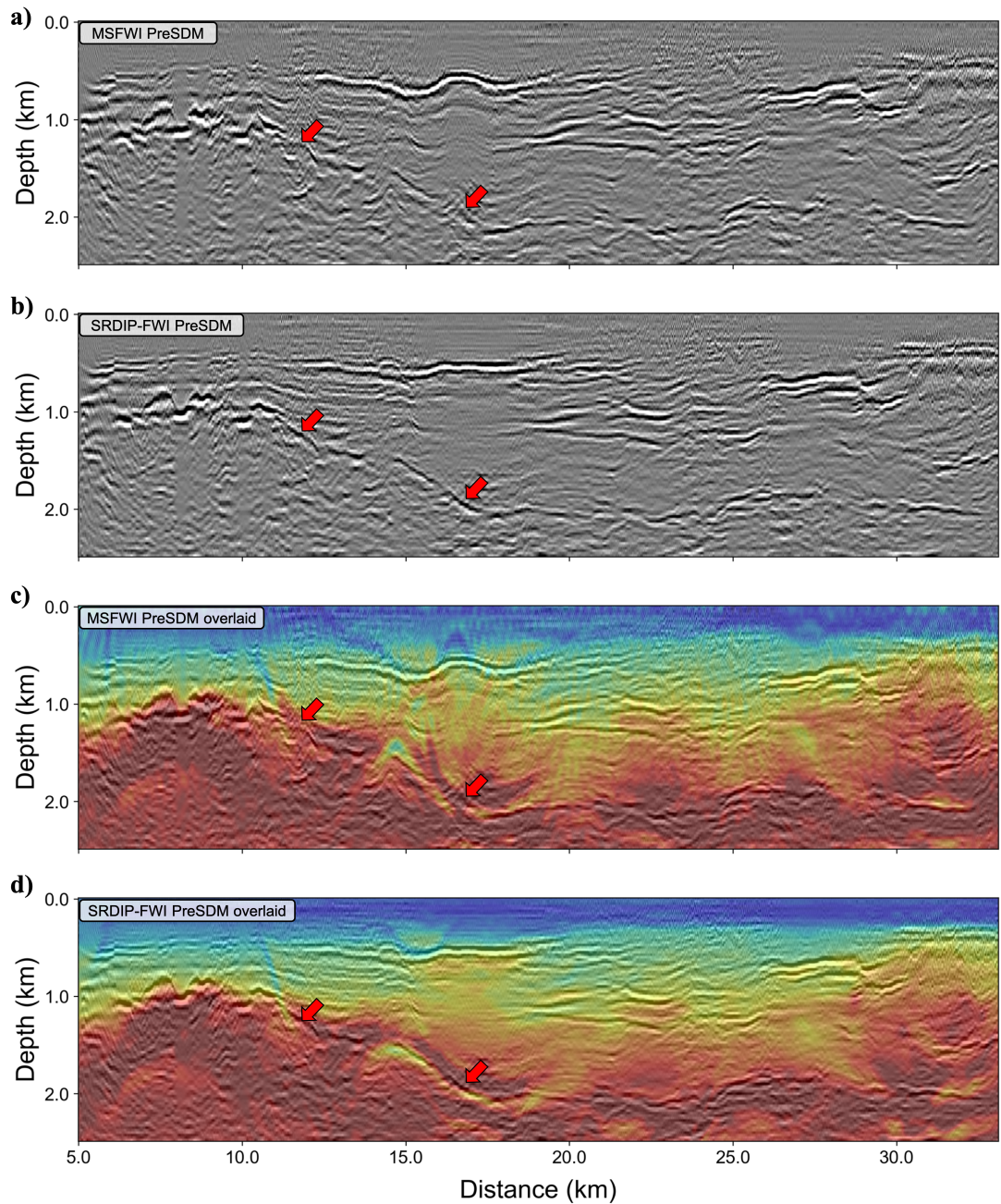

**Figure 10.** PreSDM results based on the velocity models obtained by MSFWI and SRDIP-FWI. (a) PreSDM result using the velocity model obtained by MSFWI; (b) PreSDM result using the velocity model obtained by SRDIP-FWI; (c) overlaid PreSDM image and the corresponding velocity mode obtained by MSFWI; (d) overlaid PreSDM image and the corresponding velocity mode obtained by SRDIP-FWI.

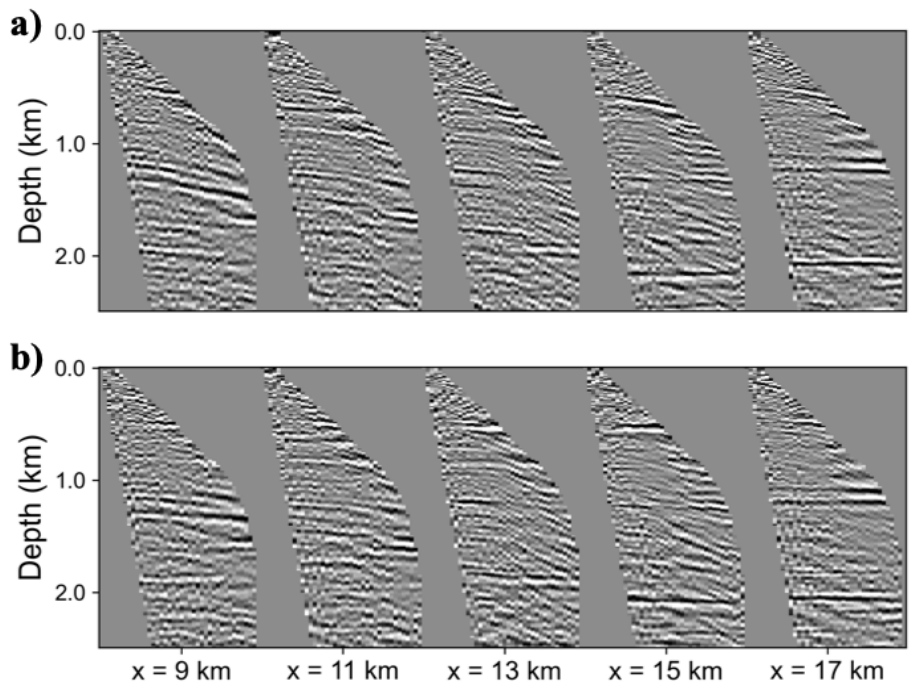

**Figure 11.** Offset-domain common image gathers (ODCIGs) based on the velocity models by MSFWI and SRDIP-FWI. (a) PreSDM ODCIGs based on the velocity model obtained by MSFWI, and (b) PreSDM ODCIGs based on the velocity model obtained by SRDIP-FWI. The horizontal positions of the five ODCIGs are at x = 9.0, 11.0, 13.0, 15.0, and 17.0 km, respectively, with a maximum offset of 2.0 km for each ODCIG.

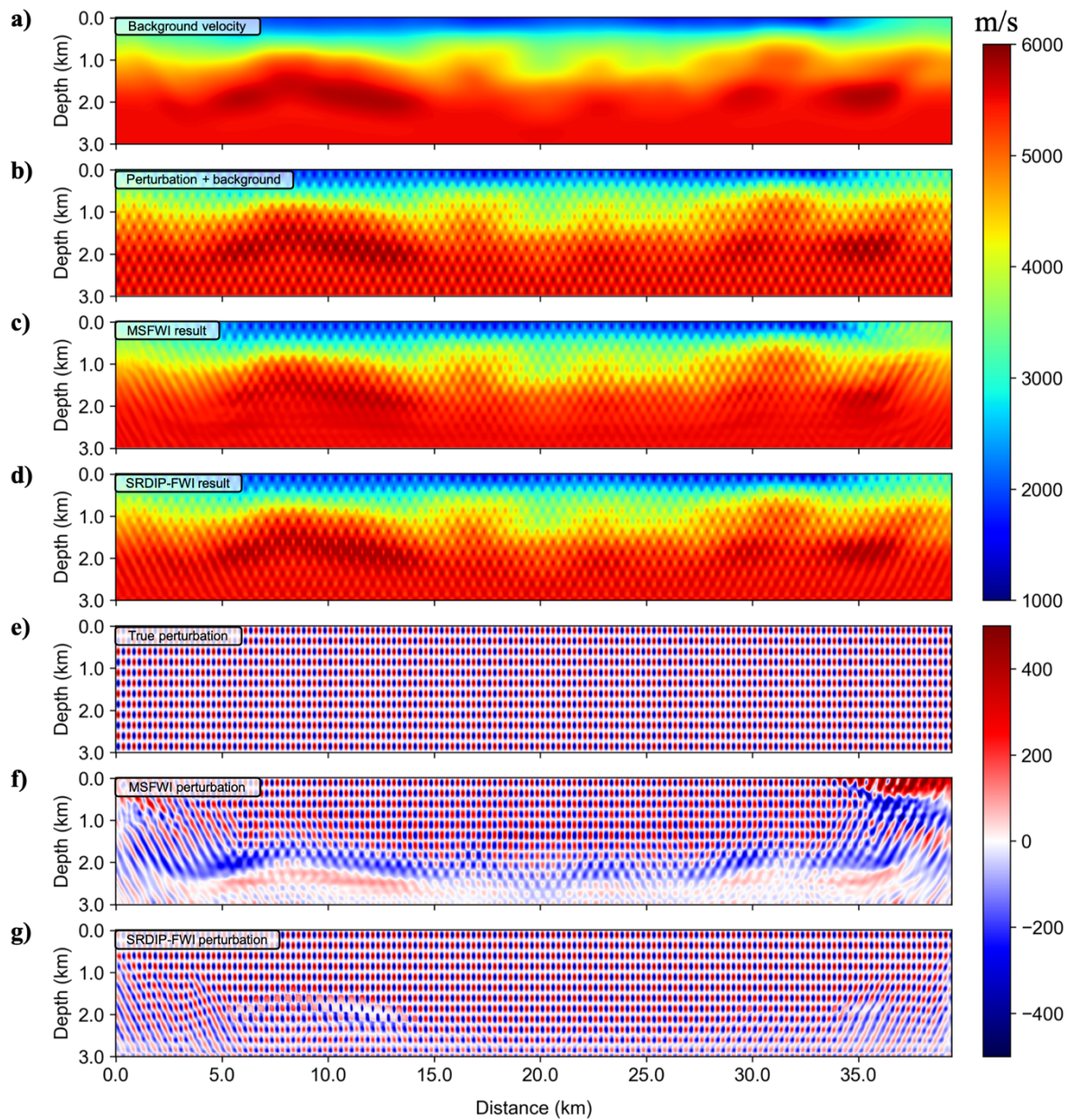

**Figure 12.** Checkerboard test for the field data with perturbations of 500 × 500 m in dimension and ±500 m/s in amplitude. (a) Background velocity model (same as Figure 7a); (b) checkerboard model constructed by adding perturbations to the background model; (c) inversion result by MSFWI; (d) inversion result by SRDIP-FWI; (e) true checkerboard perturbations; (f) recovered perturbation by MSFWI; (g) recovered perturbation by SRDIP-FWI.

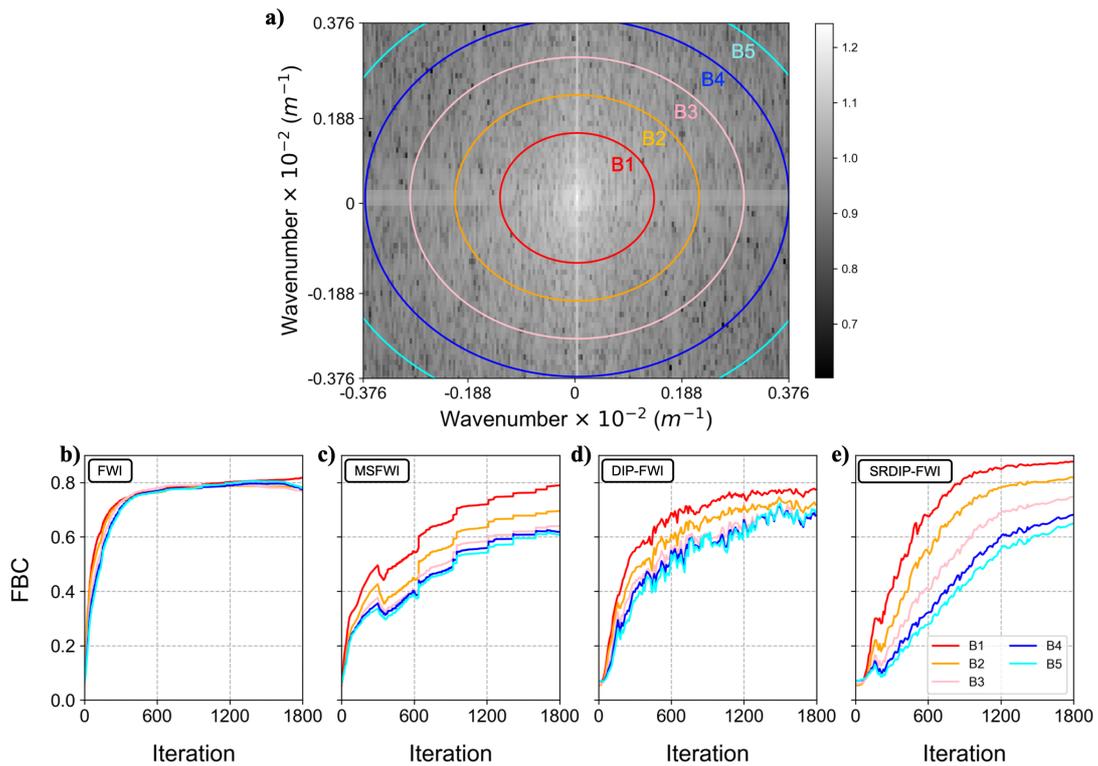

**Figure 13.** BP salt model in the wavenumber domain and the FBC curves with different FWI approaches. (a) The wavenumber domain representation of the BP salt model shown in Figure 2a. The domain is divided into five concentric bands (B1 – B5) from low to high wavenumbers. (b), (c), (d) and (e) show the FBC curves with iterations for FWI, MSFWI, DIP-FWI and SRDIP-FWI, respectively.

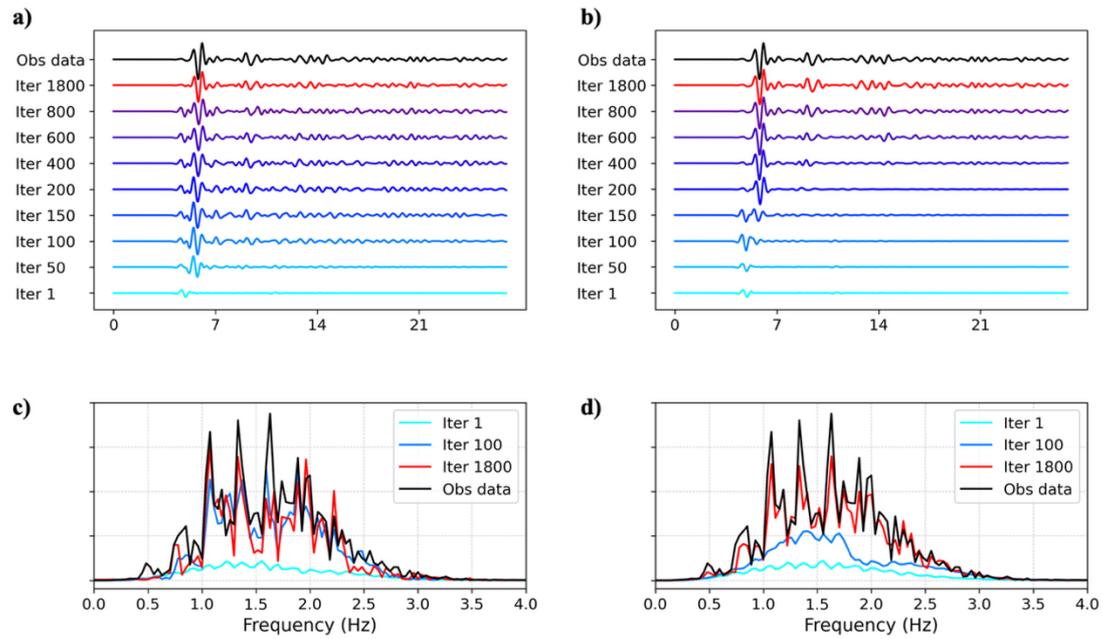

**Figure 14.** Comparison of a single trace matching based on the BP salt models inverted by FWI and SRDIP-FWI. (a) Comparison between the "observed" and synthetic traces calculated with the inverted velocity models at different iterations with FWI; (b) similar to (a), but for SRDIP-FWI; (c) and (d) show the spectra of the waveforms at selected iterations in (a) and (b), respectively.

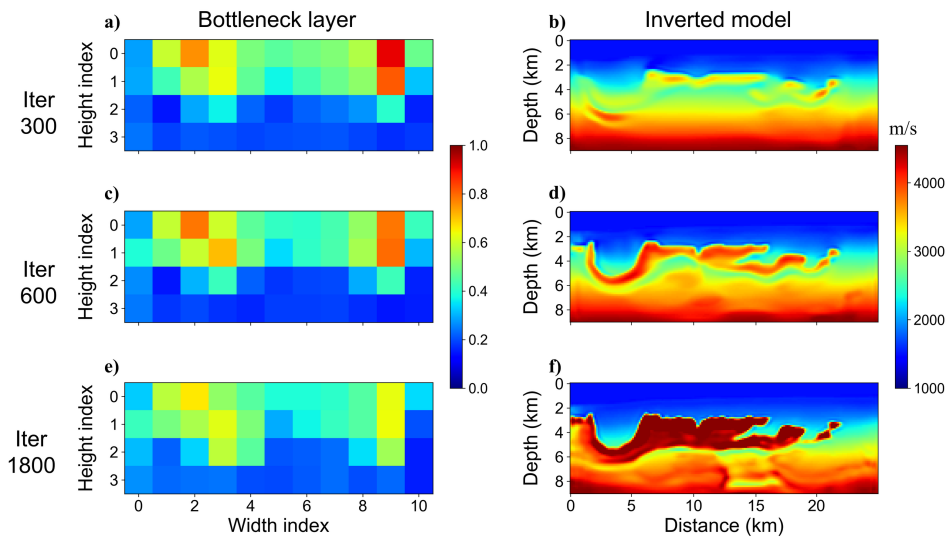

**Figure 15.** Feature maps extracted from the bottleneck layer of the reparameterizing DIP network and corresponding inversion results. (a) and (b) show the feature map and inversion result at iteration 300, (c) and (d) at iteration 600, and (e) and (f) at iteration 1800, respectively.

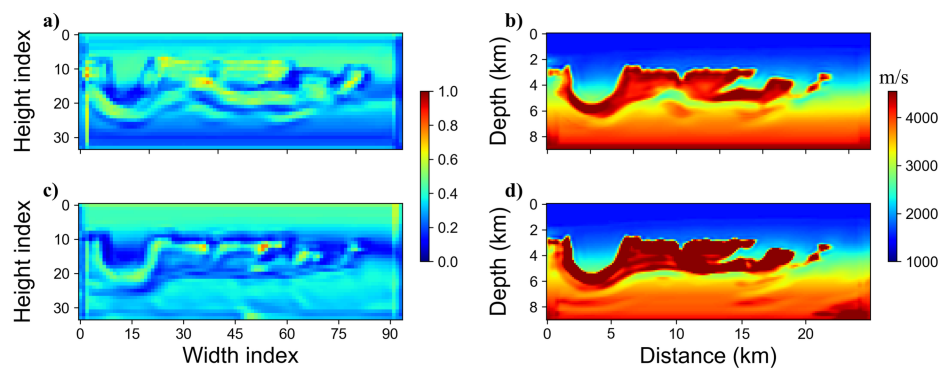

**Figure 16.** Ablation tests for the depth and number of parameters of the U-Net used in the reparameterizing DIP network in SRDIP-FWI. (a) The bottleneck feature map of the single-level U-Net；(b) the corresponding inverted model using the single-level raparameterizing DIP network; (c) and (d) are similar to (a) and (b), but for an expanded single-level U-Net with roughly the same number of parameters as the four-

level U-Net (Figure 15).

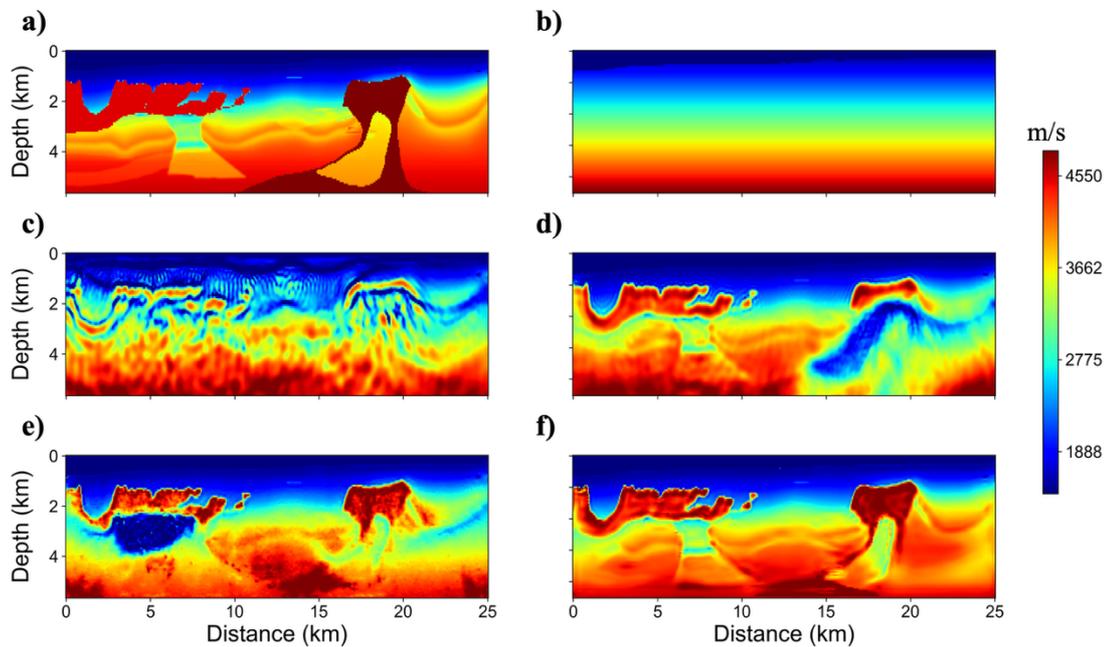

**Figure A-1.** Inversion results for the full BP model with different FWI approaches. (a) The true full BP model; (b) the initial 1D model; inversion results using (c) FWI, (d) MSFWI, (e) DIP-FWI, and (f) SRDIP-FWI.

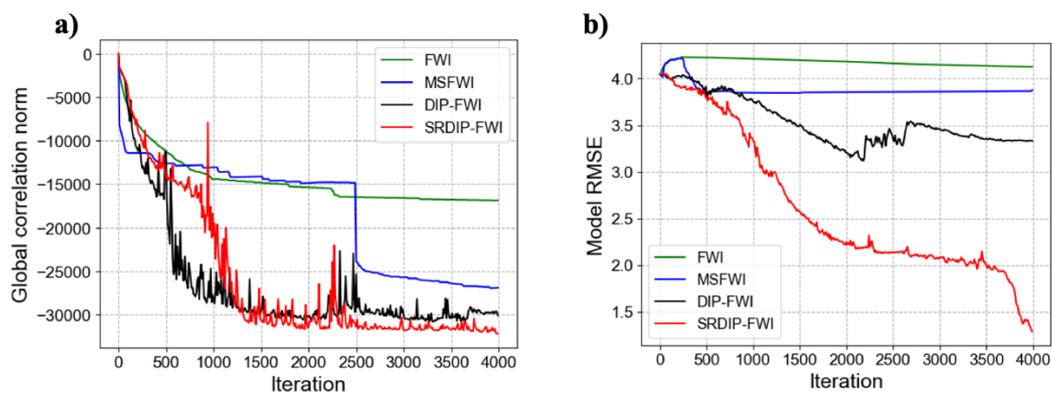

**Figure A-2.** Data global correlation norm and model RMSE with iterations for the inverted full BP model using FWI, MSFWI, DIP-FWI, and SRDIP-FWI. (a) Global correlation norm curves with iterations, and (b) model RMSE curves with iterations.

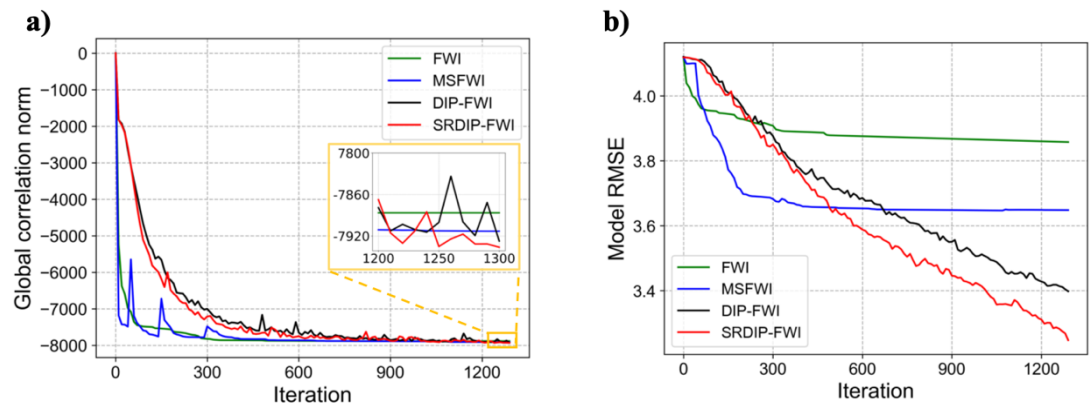

**Figure C-1.** Data global correlation norm and model RMSE with iterations for the Marmousi model using FWI, MSFWI, DIP-FWI, and SRDIP-FWI. (a) Global correlation norm curves with iterations, and (b) model RMSE curves with iterations.

**Algorithm 1** Self-reinforced reparameterized full waveform inversion (SRDIP-FWI)

**Input:** Observed data $d_{obs}$, initial model $m_0$, network parameters learning rate $\eta_\theta$, network input learning rate $\eta_z$, maximum iterations $K$, random noise input $z_r^0$.

**Output:** Optimized network parameters $\theta$, inverted model $m_K$.

1: Define network architecture $f_\theta(z_r)$, initialize network parameters $\theta \leftarrow \theta_0$ and initial input $z_r \leftarrow z_r^0$;

2: Set maximum iterations $K$;

3: **for** i $= 0,1,2,\dots,K$ do

4: $\qquad \boldsymbol{m_i} = m_0 + f_{\theta^i}(z_r^i)$ ;

5: $\qquad$ Forward modeling $\boldsymbol{d_{syn}} \leftarrow \mathcal{L}(m_i)$;

6: $\qquad$ **Step I: Update network parameters**

7: $\qquad\qquad$ Fix network input $z_r^i$;

9: $\qquad\qquad$ Compute FWI misfit $\boldsymbol{\phi(m)} \leftarrow \mathcal{D}(d_{obs}, d_{syn})$;

10: $\qquad\qquad$ Compute FWI gradient $\boldsymbol{\partial\phi(m)/\partial m}$ using adjoint method;

11: $\qquad\qquad$ Compute the $\boldsymbol{\partial\phi(m)/\partial\theta_i}$;

12: $\qquad\qquad$ Update network parameters $\boldsymbol{\theta_i}$ using Adam optimization with $\eta_\theta$

13: $\qquad\qquad\qquad \boldsymbol{\theta^{i+1}} = \theta^i - \eta_\theta \cdot \partial\phi(m)/\partial\theta_i$;

14: $\qquad\qquad$ Generate intermediate model $\widetilde{m}_{i+1} \leftarrow m_0 + f_{\theta^{i+1}}(z_r^i)$ ;

15: $\qquad$ **Step II: Update network input**

16: $\qquad\qquad$ Fix network parameters $\theta^{i+1}$ ;

17: $\qquad\qquad$ Compute steering loss $\mathcal{J}(z_r^i) = \left\| z_r^i - \widetilde{m}^{i+1} \right\|$ ;

18: $\qquad\qquad$ Update network input $\boldsymbol{z_r^i}$ using Adam optimization with $\eta_z$;

19: $\qquad\qquad\qquad \boldsymbol{z_r^{i+1}} = z_r^i - \eta_z(z_r^i - m^{i+1})$

20: $\qquad\qquad$ Update model $\boldsymbol{m_{i+1}} \leftarrow m_0 + f_{\theta^{i+1}}(z_r^{i+1})$ ;

21: **end for**

22: **return** $\theta_K$, $m_K$

**Table 1. Parameters for MSFWI in inverting for two different models.**

| BP model (part) | Stage 1 | Stage 2 | Stage 3 | Stage 4 | Stage 5 |
|---|---|---|---|---|---|
| Frequency (Hz) | < 0.6 | < 1.2 | < 1.8 | < 3.0 | < 5 |
| Time window (s) | 2.5 | 5 | 10 | 15 | 27 |
| Iterations | 300 | 300 | 300 | 300 | 600 |
| **Marmousi2 model** | Stage 1 | Stage 2 | Stage 3 | Stage 4 | Stage 5 |
| Frequency (Hz) | < 3 | < 5 | < 8 | < 15 | < 15 |
| Time window (s) | 0.5 | 1 | 2 | 4 | 4.8 |
| Iterations | 50 | 100 | 150 | 200 | 800 |
| **BP model (full)** | Stage 1 | Stage 2 | Stage 3 | Stage 4 | Stage 5 |
| Frequency (Hz) | < 0.6 | < 1.2 | < 1.8 | < 5 | < 10 |
| Time window (s) | 2.5 | 5 | 10 | 15 | 20 |
| Iterations | 250 | 250 | 250 | 250 | 3000 |

**Table 2. Model metrics of FWI, MSFWI, DIP-FWI, and SRDIP-FWI for part of the BP model. (Smaller RMSE or larger SSIM/PSNR indicate better model quality)**

|  | RMSE↓ | SSIM↑ | PSNR↑ |
|---|---|---|---|
| FWI | 5.983 | 0.531 | 13.73 |
| MSFWI | 4.496 | 0.696 | 16.22 |
| DIP-FWI | 6.486 | 0.630 | 13.03 |
| SRDIP-FWI | **2.109** | **0.869** | **22.79** |

**Table 3. MSFWI parameters used in different inversion stages for the field data acquired in Inner Mongolia, China.**

|  | Stage 1 | Stage 2 | Stage 3 | Stage 4 | Stage 5 |
|---|---|---|---|---|---|
| Frequency (Hz) | 4~8 | 4~8 | 4~16 | 4~16 | 4~16 |
| Time window (s) | 0.1 | 0.25 | 0.1 | 0.25 | 0.4 |
| Iterations | 20 | 20 | 20 | 20 | 20 |

**Table A-1. Model metrics for the inverted full BP model using FWI, MSFWI, DIP-FWI, and SRDIP-FWI (lower RMSE and higher SSIM/PSNR indicate better model quality)**

|  | RMSE↓ | SSIM↑ | PSNR↑ |
|---|---|---|---|
| FWI | 4.129 | 0.550 | 15.67 |
| MSFWI | 3.075 | 0.771 | 16.22 |
| DIP-FWI | 3.342 | 0.730 | 17.50 |
| SRDIP-FWI | **1.287** | **0.856** | **25.80** |

**Table C-1. Model metrics of FWI, MSFWI, DIP-FWI, and SRDIP-FWI for the Marmousi model. (Smaller RMSE and larger SSIM/PSNR indicate better model quality)**

|  | RMSE↓ | SSIM↑ | PSNR↑ |
|---|---|---|---|
| FWI | 3.871 | 0.522 | 15.86 |
| MSFWI | 3.648 | 0.576 | 16.37 |
| DIP-FWI | 3.392 | 0.584 | 17.00 |
| SRDIP-FWI | **3.255** | **0.608** | **17.36** |